\documentclass[12pt]{article}
\usepackage{times}
\usepackage{a4}
\usepackage{latexsym}
\usepackage{amsmath}
\usepackage{amsfonts}
\usepackage{graphicx}
\usepackage{fancyhdr,lastpage}
\newcommand{\rank}{\mathrm{r}}
\newcommand{\sgn}{\mathrm{sgn}}
\newcommand{\tr}{\mathrm{tr}}

\newcommand{\R}{\mathbb{R}}

\newtheorem{theorem}{Theorem}
\newtheorem{corollary}[theorem]{Corollary}
\newtheorem{definition}{Definition}
\newenvironment{proof}[1][Proof]{\noindent\textbf{#1.} }{\ \rule{0.5em}{0.5em}}
\begin{document}

\title{Random Matrix Theory and \\
Robust Covariance Matrix Estimation \\
for Financial Data}
\author{Gabriel Frahm\thanks{Email: {\tt frahm@ccrl-nece.de}.}\hspace{.2cm} \& Uwe Jaekel\thanks{Email: {\tt jaekel@ccrl-nece.de}.}\\[.5cm]
C\&C Research Laboratories, NEC Europe Ltd.\\
Rathausallee 10, 53757 Sankt Augustin, Germany}
\maketitle

\vspace{-1cm}
\begin{abstract}
The traditional class of elliptical distributions is extended to allow for asymmetries. A completely robust dispersion matrix estimator (the `spectral estimator') for the new class of `generalized elliptical distributions' is presented. It is shown that the spectral estimator corresponds to an M-estimator proposed by Tyler (1983) in the context of elliptical distributions. Both the generalization of elliptical distributions and the development of a robust dispersion matrix estimator are motivated by the stylized facts of empirical finance. Random matrix theory is used for analyzing the linear dependence structure of high-dimensional data. It is shown that the Mar\v{c}enko-Pastur law fails if the sample covariance matrix is considered as a random matrix in the context of elliptically distributed and heavy tailed data. But substituting the sample covariance matrix by the spectral estimator resolves the problem and the Mar\v{c}enko-Pastur law remains valid.
\end{abstract}

\section{Motivation}

Short-term financial data usually exhibit similar properties called `stylized facts' like, e.g., leptokurtosis,
dependence of simultaneous extremes, radial asymmetry, vola\-tility clustering, etc., especially if the
log-price changes (called the `log-returns') of stocks, stock indices, and foreign exchange rates are
considered. Particularly, high-frequency data usually are non-stationary, have jumps, and are strongly
dependent. Cf., e.g., Bouchaud, Cont, and Potters, 1998, Breymann, Dias, and Embrechts, 2003, Eberlein and
Keller, 1995, Embrechts, Frey, and McNeil, 2004 (Section 4.1.1), Engle, 1982, Fama, 1965, Junker and May, 2002, Mandelbrot, 1963, and Mikosch, 2003 (Chapter 1).

Figure 1 contains QQ-plots of $\text{GARCH}(1,1)$ residuals of daily log-returns of the NASDAQ and the
S\&P 500 indices from 1993-01-01 to 2000-06-30. It is clearly indicated that the normal distribution hypothesis is not appropriate for the loss parts of the distributions whereas the Gaussian law seems to be acceptable for the profit parts. Hence the probability of extreme losses is higher than suggested by the normal distribution assumption.

\begin{center}
\includegraphics[scale=.34]{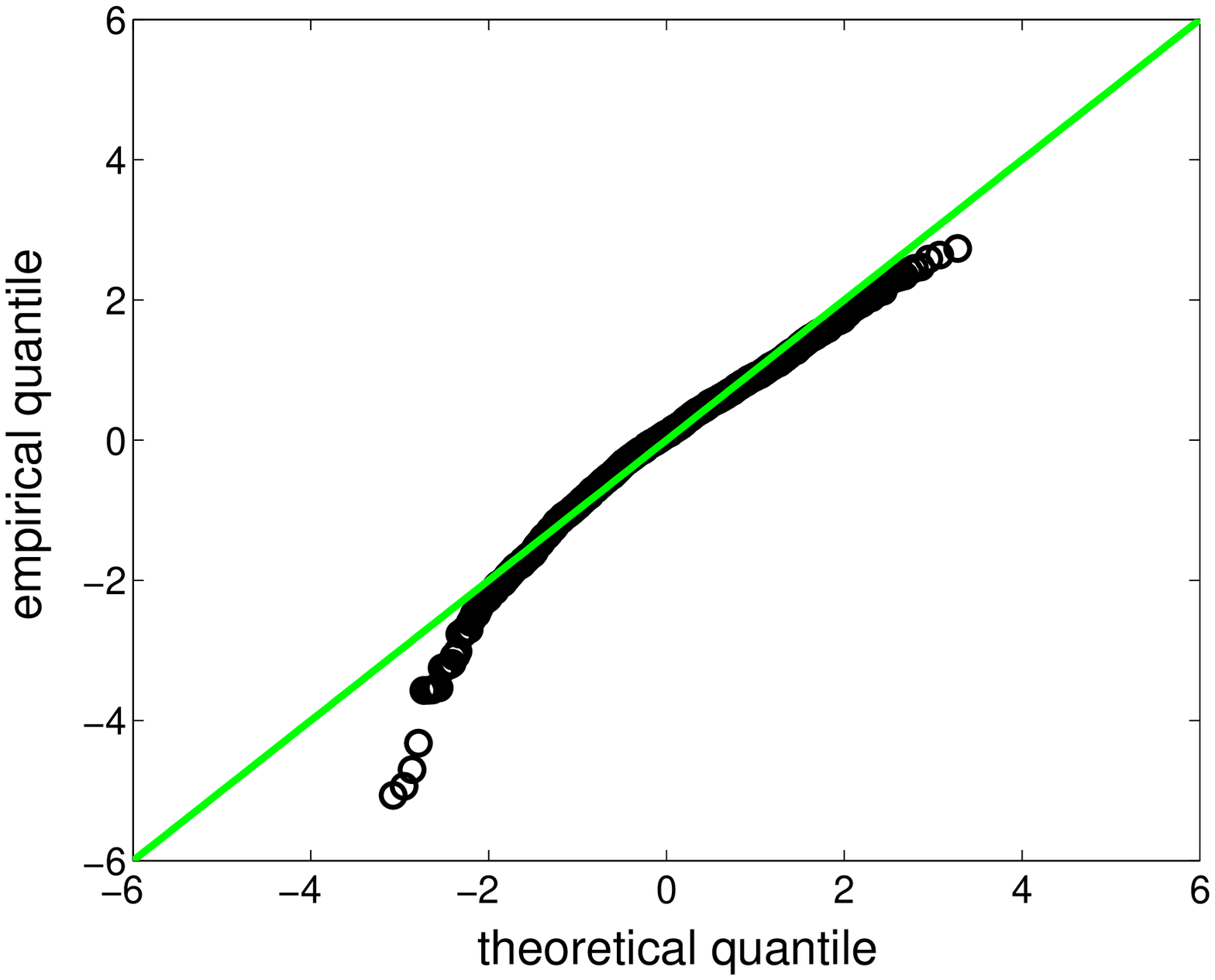}
\includegraphics[scale=.34]{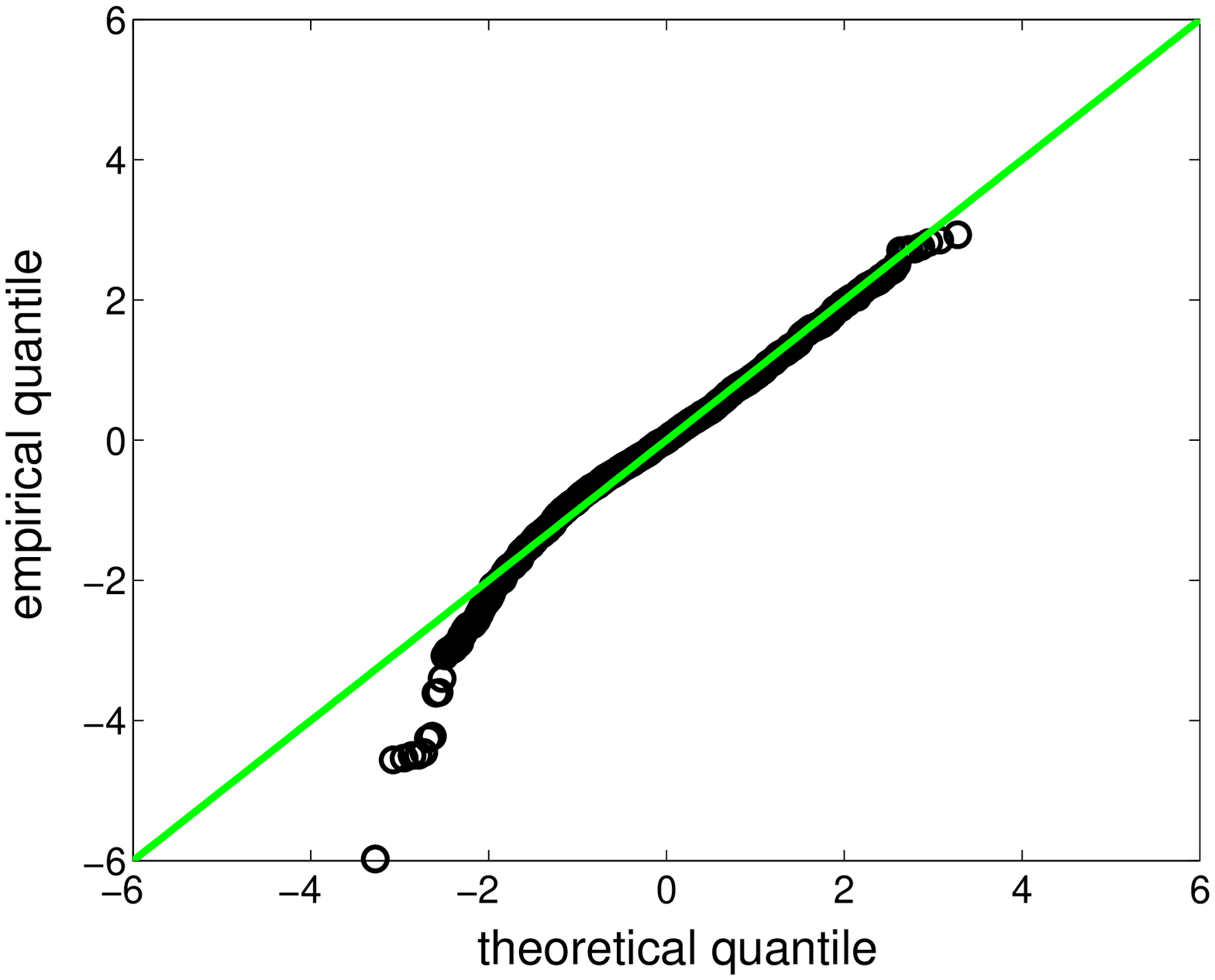}\\[.25cm]
\end{center}
{\bf Fig. 1:} QQ-plots of NASDAQ (left hand) and S\&P 500 (right hand) $\text{GARCH}(1,1)$ residuals from
1993-01-01 to 2000-06-30 ($n=1892$).\\[.25cm]

The next picture shows the joint distribution of the GARCH residuals considered above.

\begin{center}
\includegraphics[scale=.35]{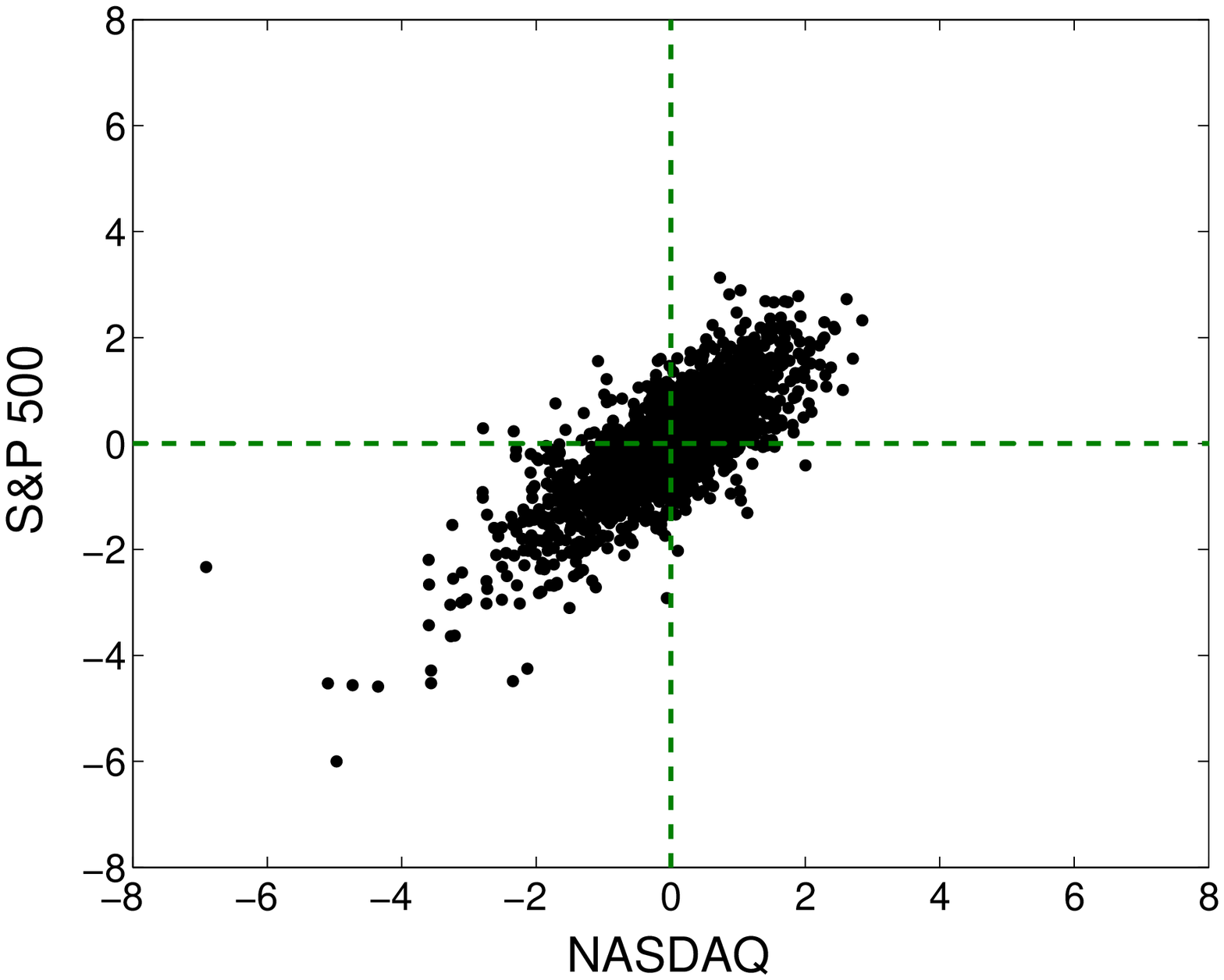}\\[.25cm]
\end{center}
{\bf Fig. 2:} NASDAQ vs. S\&P 500 $\text{GARCH}(1,1)$ residuals from 1993-01-01 to 2000-06-30 ($n=1892$).\\[.25cm]

Except for one element all extremes occur simultaneously. The effect of simultaneous extremes can be observed
more precisely in the following picture. It shows the total numbers of S\&P 500 stocks whose absolute values of
daily log-returns exceeded $10\%$ for each trading day during 1980-01-02 to 2003-11-26. On the 19th October
1987 (i.e. the `Black Monday') there occurred 239 extremes. This is suppressed for the sake of transparency.

\begin{center}
\includegraphics[scale=.35]{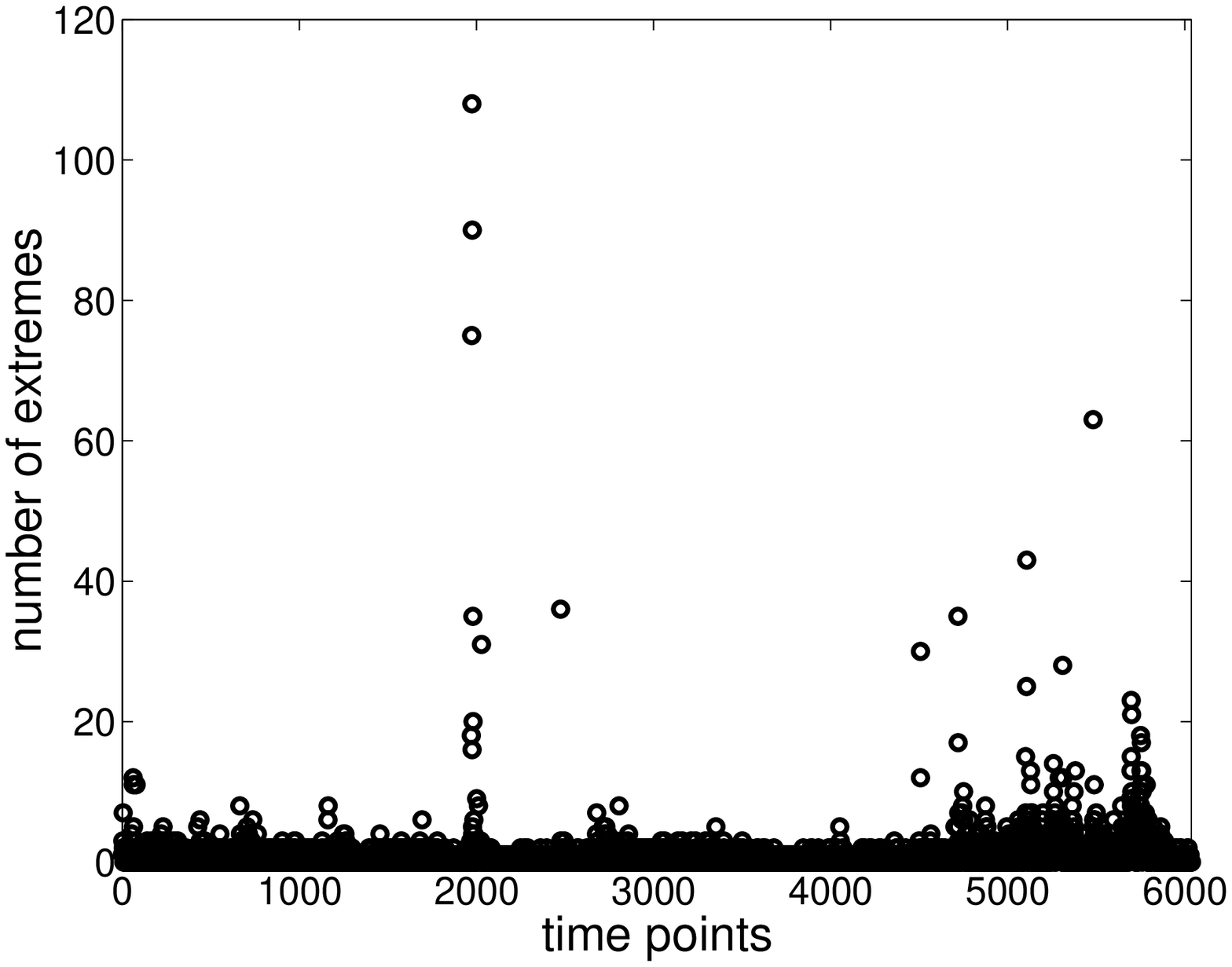}\\[.25cm]
{\bf Fig. 3:} Number of extremes in the S\&P 500 during 1980-01-02 to 2003-11-26.\\[.25cm]
\end{center}

The latter figure shows the concomitance of extremes. If extremes would occur independently then the number of
extremal events (no matter if losses or profits) should be small and all but constant over time. Obviously,
this is not the case. In contrast one can see the October Crash of 1987 and several extremes which occur
permanently since the beginning of the bear market in 2000. Hence there is an increasing tendency of
simultaneous losses which is probably due to globalization effects and relaxed market regulation. The phenomenon of simultaneous extremes is often denoted by `asymptotic dependence' or `tail dependence'.

The traditional class of elliptically symmetric distributions (Cambanis, Huang, and Simons, 1981, Fang, Kotz,
and Ng, 1990, and Kelker, 1970) is often proposed for the modeling of financial data (cf., e.g., Bingham and
Kiesel, 2002). But elliptical distributions suffer from the pro\-perty of radial symmetry. The pictures above
show that financial data are not always symmetrically distributed. For this reason the authors will bear on the
assumption of gene\-ralized elliptically distributed (Frahm, 2004) log-returns. This allows for the modeling of tail dependence and radial asymmetry.

The quintessence of modern portfolio theory is that the portfolio diversification effect depends essentially on
the covariances. But the parameters for portfolio optimization, i.e. the mean vector and the covariance matrix,
have to be estimated. Especially for portfolio risk minimization a reliable estimate of the covariance matrix
is necessary (Chopra and Ziemba, 1993). For covariance matrix estimation generally one should use as much
available data as possible. But since daily log-returns and all the more high-frequency data are not normally
distributed, standard estimators like the sample covariance matrix may be highly inefficient leading to
erroneous implications (see, e.g., Oja, 2003 and Visuri, 2001). This is because the sample covariance matrix is
very sensitive to outliers. The smaller the distribution's tail index (Hult and Lindskog, 2002), i.e. the
heavier the tails of the log-return distributions the higher the estimator's variance. So the quality of the
parameter estimates depends essentially on the true multivariate distribution of log-returns.

In the following it is shown how the linear dependence structure of generalized elliptical random vectors can be
estimated robustly. More precisely, it is shown that Tyler's (1987) robust M-estimator for the dispersion matrix
$\Sigma$ of elliptically distributed random vectors remains completely robust for generalized elliptically
distributed random vectors. This estimator is not disturbed neither by asymmetries nor by outliers and all the
available data points can be used for estimation purposes. Further, the impact of high-dimensional (financial)
data on statistical inference will be discussed. This is done by referring to a branch of statistical physics 
called `Random Matrix Theory' (Hiai and Petz, 2000 and Mehta, 1990). Random matrix theory (RMT) is concerned
with the distribution of eigenvalues of high-dimensional randomly generated matrices. If each component of a sample is independent and identically distributed then the distribution of the eigenvalues of the sample covariance matrix converges to a specified law which does not depend on the specific distribution of the sample components. The circumstances under which this result of RMT can be properly adopted to generalized elliptically distributed data will be examined.

\section{Generalized Elliptical Distributions}

It is well known that an elliptically distributed random vector $X$ can be represented stochastically by
$X\! =_{\mathrm{d}}\! \mu +\mathcal{R}\Lambda U^{\left( k\right)}$, where $\mu\in\R^{d}$, $\Lambda\in\R^{d\times k}$
with $\rank(\Lambda)=k$, $U^{\left( k\right) }$ is a $k$-dimensional random vector uniformly distributed on the
unit hypersphere $\mathcal{S}^{k-1}$, and $\mathcal{R}$ is a nonnegative random variable stochastically
independent of $U^{\left( k\right) }$. The positive semi-definite matrix $\Sigma := \Lambda\Lambda^{\mathrm{T}}$ characterizes the linear dependence structure of $X$ and is referred to as the `dispersion matrix'.

\begin{definition}[Generalized elliptical distribution]
The $d$-dimensional random vector $X$ is said to be `generalized elliptically distributed' if and only if

\begin{equation*}
X\overset{\mathrm{d}}{=}\mu +\mathcal{R}\Lambda U^{\left( k\right) }.
\end{equation*}

where $U^{\left( k\right) }$ is a $k$-dimensional random vector uniformly distributed on $\mathcal{S}^{k-1}$,
$\mathcal{R}$ is a random variable, $\mu \in \R^{d}$, and $\Lambda \in \R^{d\times k}$.

\end{definition}

Note that the definition of generalized elliptical distributions preserves all the ordinary components of
elliptically symmetric distributions (i.e. $\mu$, $\Sigma$, and $\mathcal{R}$). But in contrast the generating
variate $\mathcal{R}$ may be negative and even more it may depend on $U^{\left( k\right) }$. It is worth to point out that the class of generalized elliptical distributions contains the class of skew-elliptical distributions (Branco and Dey, 2001, and Frahm, 2004, Section 3.2).

The next figure shows once again the joint distribution of the GARCH residuals of the NASDAQ and
S\&P 500 log-returns from 1993-01-01 to 2000-06-30 from Figure 2. The right hand of Figure 4 contains
simulated GARCH residuals on the basis of a generalized $t$-distribution. More precisely, the generating variate
$\mathcal{R}$ corres\-ponds to $\sqrt{\nu \cdot \chi _{2}^{2}/\chi _{\nu }^{2}}\,$ but the number of degrees of
freedom $\nu$ depends on $U^{(2)}$, i.e. $\nu = 4 + 996\cdot\left(\delta(\Lambda u / \|\Lambda
u\|_{2},v\right))^{3}$ $(\|u\|_{2}=1)$. Here $\delta$ is a function that measures the distance between $\Lambda
u / \|\Lambda u\|_{2}$ and the reference vector $v=\left(-\cos \left( \pi /4\right) ,-\sin \left( \pi /4\right)
\right)$, $\delta(u,v) := \angle(u,v)/\pi = \arccos(u^{\mathrm{T}} v)/\pi$. Hence, random vectors which are close to the
reference vector (i.e. close to the `perfect loss scenario') are supposed to be $t$-distributed with $\nu=4$
degrees of freedom whereas random vectors which are opposite are assumed to be nearly Gaussian ($\nu=1000$)
distributed. This is consistent with the phenomenon observed in Figure 1. The pseudo-correlation coefficient is
set to $0.78$.

\begin{center}
\includegraphics[scale=.34]{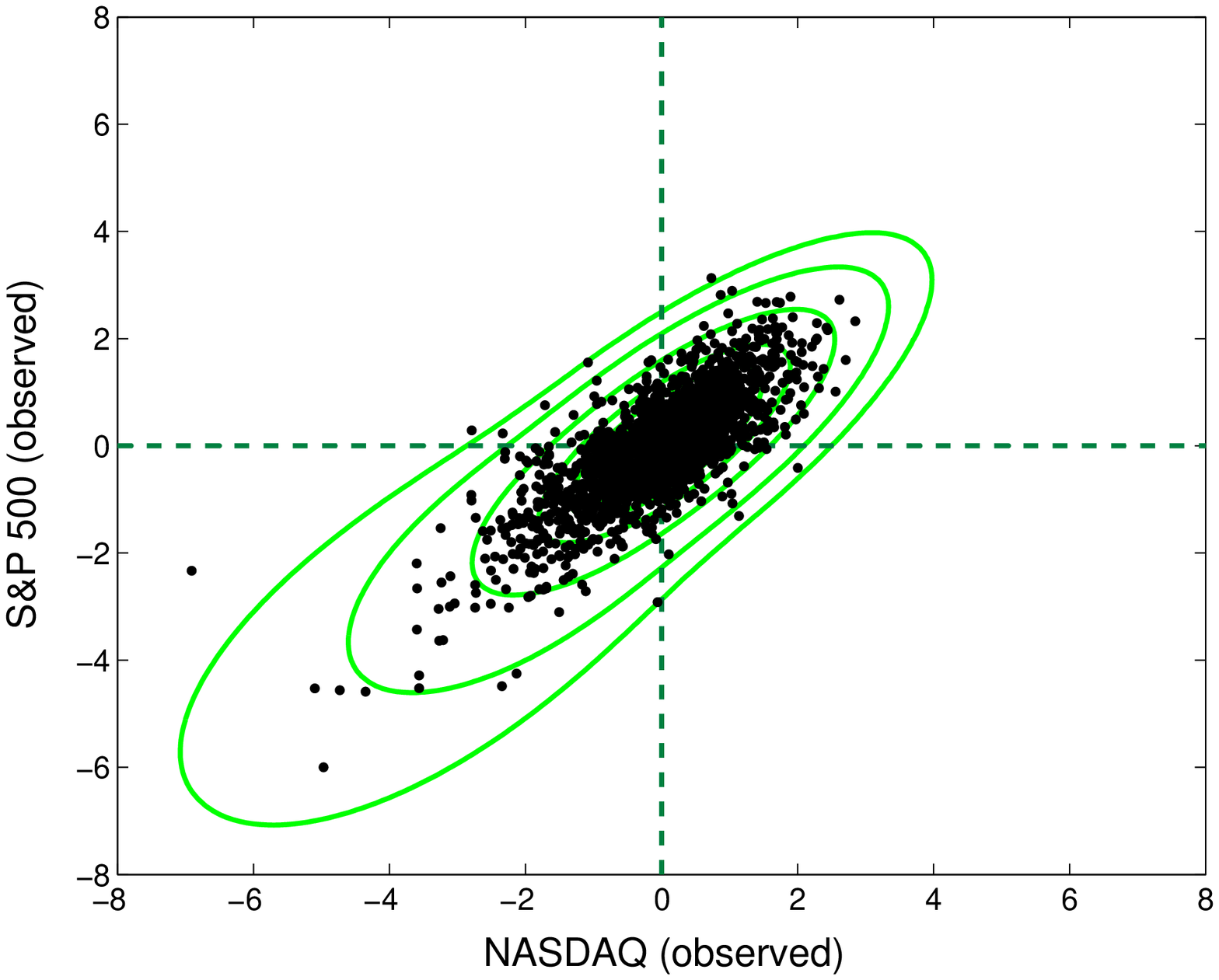}
\includegraphics[scale=.34]{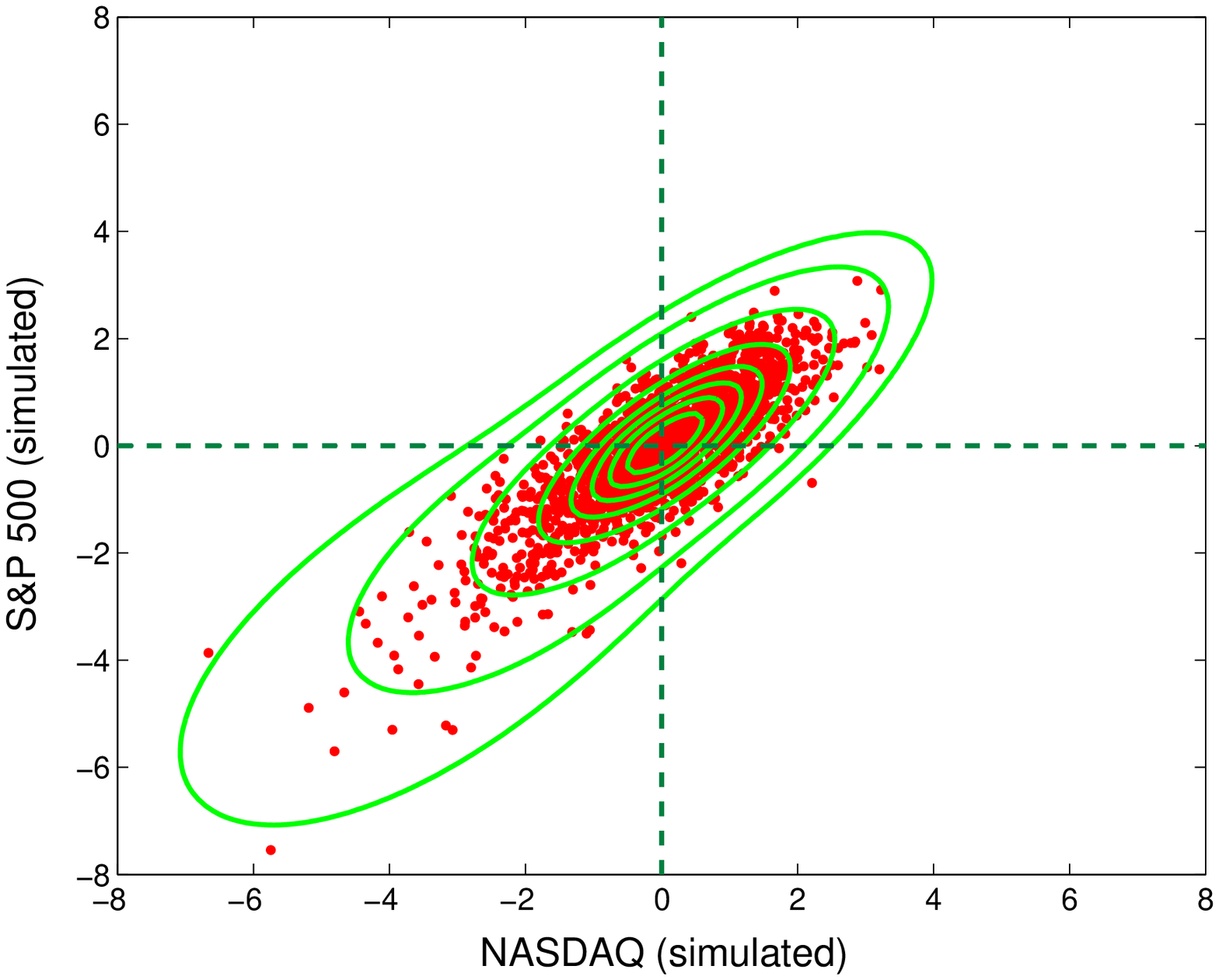}\\[.25cm]
\end{center}
{\bf Fig. 4:} Observed $\text{GARCH}(1,1)$ residuals of NASDAQ and S\&P 500 (left hand) and simulated
generalized $t$-distributed random noise ($n=1892$) (right hand).\\[.25cm]

\section{Robust Covariance Matrix Estimation}

It is well-known that the sample covariance matrix corresponds both to the moment estimator and to the
ML-estimator for the dispersion matrix $\Sigma$ of normally distributed data. But given any other elliptical
distribution family the dispersion matrix usually does not correspond to the covariance matrix. Generally,
robust covariance matrix estimation means to estimate the dispersion matrix, that is the covariance matrix up
to a scaling constant. There are many applications like, e.g., principal components analysis, canonical
correlation analysis, linear discriminant ana\-lysis, and multivariate regression where only the dispersion
matrix is demanded (Oja, 2003). Particularly, by Tobin's two-fund separation theorem (Tobin, 1958) the
optimal portfolio of risky assets does not depend on the scale of the covariance matrix. Thus in the following
we will loosely speak of `covariance matrix estimation' rather than of estimating the dispersion matrix for the
sake of simplicity.

As mentioned before the true linear dependence structure of elliptically distributed data can not be estimated efficiently by the sample covariance matrix, generally. Especially, if the data stem from a regularly varying random
vector the smaller the tail index, i.e. the heavier the tails the larger the estimator's variance. But in the following it is shown that there exists a completely robust alternative to the sample covariance matrix.

Let $X$ be a $d$-dimensional generalized elliptically distributed random vector where $\mu$ is supposed to be
known, $\Lambda \in \R^{d\times k}$ with $\rank(\Lambda)=d$, and $P(\mathcal{R}=0)=0$. Further, let the unit
random vector generated by $\Lambda$ be defined as

\begin{equation*}
S := \frac{\Lambda U^{\left( k\right) }}{ {\big |\!|}\Lambda U^{\left( k\right) }{\big |\!|}_{2}}.
\end{equation*}

Due to the stochastic representation of $X$ the following relations hold,

\begin{equation*}
\frac{X-\mu}{{\big |\!|}X-\mu{\big |\!|}_{2}}\overset{\mathrm{d}}{=}%
\frac{\mathcal{R}\Lambda U^{\left( k\right) }}{
{\big |\!|}\mathcal{R}\Lambda U^{\left( k\right) }{\big |\!|}_{2}}\overset{\mathrm{a.s.}}{=}%
\pm\frac{\Lambda U^{\left( k\right) }}{ {\big |\!|}\Lambda U^{\left( k\right) }{\big |\!|}_{2}}=\pm S,
\end{equation*}

where $\pm :=\sgn(\mathcal{R})$. The random vector $\pm S$ does not depend on the absolute value of
$\mathcal{R}$. So it is completely robust against extreme outcomes of the generating variate. But the sign of
$\mathcal{R}$ still remains and this may depend on $U^{\left( k\right) }$, anymore. Suppose for the moment that
$\pm$ is known for each realization of $\mathcal{R}$. Then the dispersion matrix of $X$ can be estimated
robustly via maximum-likelihood estimation using the density function of $S$ which is only a function of
$\Lambda$. This is given by the next theorem.

\begin{theorem}
The spectral density function of the unit random vector generated by $\Lambda \in \R^{d\times k}$ corresponds
to

\begin{equation*}\label{spectral_density}
s\longmapsto \psi \left( s\right) =\frac{\Gamma \left( \frac{d}{2}\right) }{2\pi ^{d/2}}\cdot \sqrt{\det
(\Sigma ^{-1})}\cdot \sqrt{s^{\mathrm{T}}\Sigma ^{-1}s}^{\,-d},\qquad \forall \ s\in \mathcal{S}^{d-1},
\end{equation*}

where $\Sigma :=\Lambda \Lambda ^{\mathrm{T}}$.
\end{theorem}

\begin{proof}
See, e.g., Frahm, 2004, pp. 59-60.\hfill \medskip
\end{proof}

Since $\psi$ is a symmetric density function the sign of $\mathcal{R}$ does not matter at all. Hence the
ML-estimation approach works even if the data are skew-elliptically distributed, for instance.

The desired `spectral estimator' is given by the fixed-point equation (Frahm, 2004, Section 4.2.2)

\begin{equation*}
\widehat{\Sigma}_{\mathrm{S}}=\frac{d}{n}\cdot \sum_{j=1}^{n}\frac{s_{j}s_{j}^{\mathrm{T}}}{s_{j}^{\mathrm{T}}\widehat{\Sigma}_{\mathrm{S}}^{-1}s_{j}},
\end{equation*}

where $s_{j}:=\left(x_{j}-\mu\right)/\left({\big |\!|}x_{j}-\mu {\big |\!|}_{2}\right)$
for $j=1,...,n$. Since the solution of the fixed-point equation is only unique up to a scaling constant in the
following it is implicitly required that the upper left element of $\widehat{\Sigma}_{\mathrm{S}}$ corresponds
to $1$.

The spectral estimator $\widehat{\Sigma}_{\mathrm{S}}$ cor\-responds to Tyler's robust M-estimator (Tyler,
1983 and Tyler, 1987) for elliptical distributions, i.e.

\begin{equation*}
\widehat{\Sigma}_{\mathrm{S}}=\frac{d}{n}\cdot \sum_{j=1}^{n}\frac{\left( x_{j}-\mu \right) \left(
x_{j}-\mu \right) ^{\mathrm{T}}}{\left( x_{j}-\mu \right) ^{\mathrm{T}}\widehat{\Sigma}_{\mathrm{S}}^{-1}\left( x_{j}-\mu \right) }.
\end{equation*}

Hence Tyler's M-estimator remains completely robust within the class of generalized elliptical distributions.

The following figure shows the sample covariance matrix (left hand) of a sample with $n=1000$ observations and
$d=500$ dimensions drawn from a multivariate $t$-distribution with $\nu=4$ degrees of freedom. Note
that the tail index of the multivariate $t$-distribution corresponds to $\nu$. Each cell of the plots represents
a matrix element where the blue colored cells symbolize small numbers and the red colored cells indicate large
numbers. The true dispersion matrix is given in the middle whereas the spectral estimate is given by the right
hand.

\begin{center}
\includegraphics[height=4.5cm,width=4.5cm]{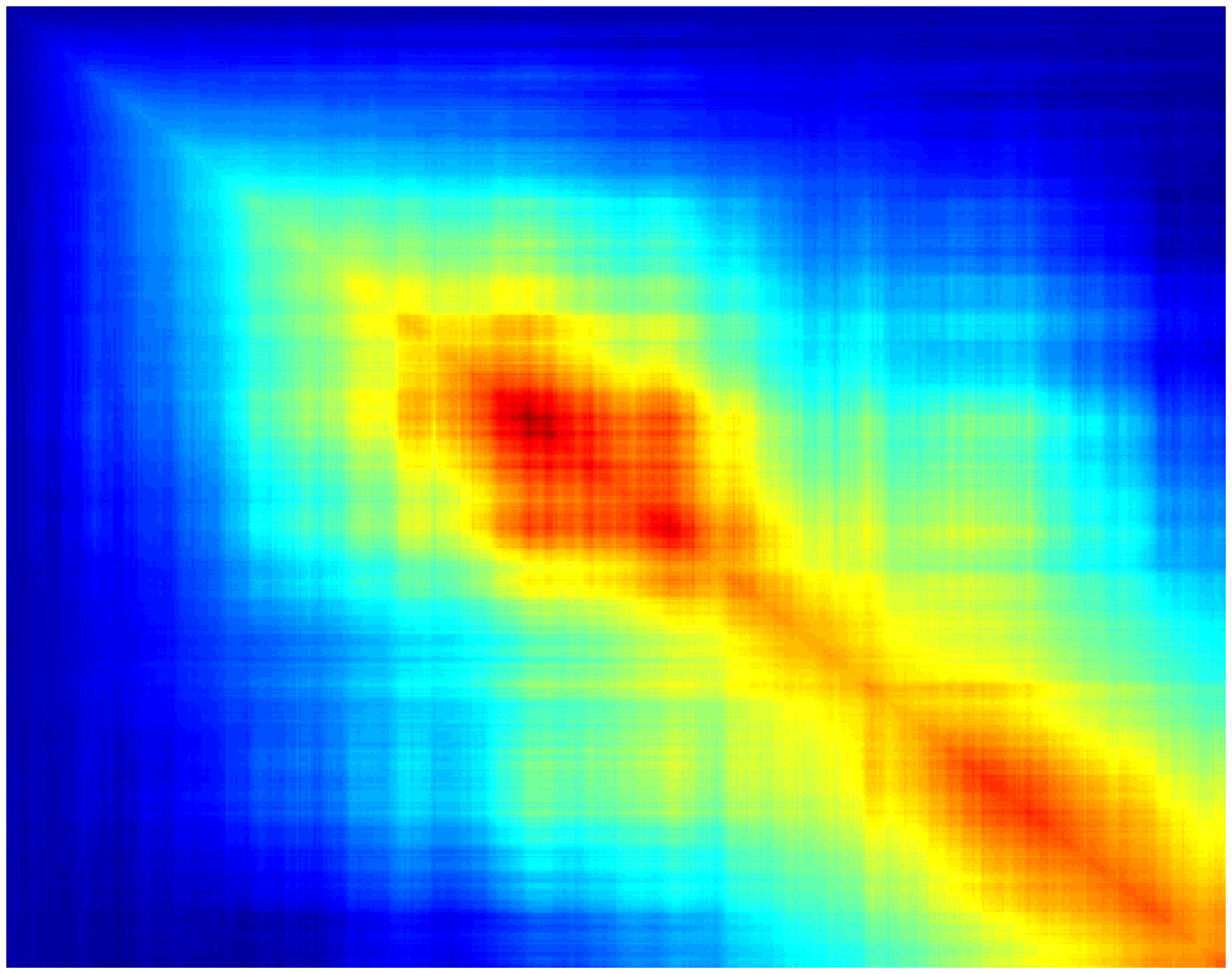}\quad
\includegraphics[height=4.5cm,width=4.5cm]{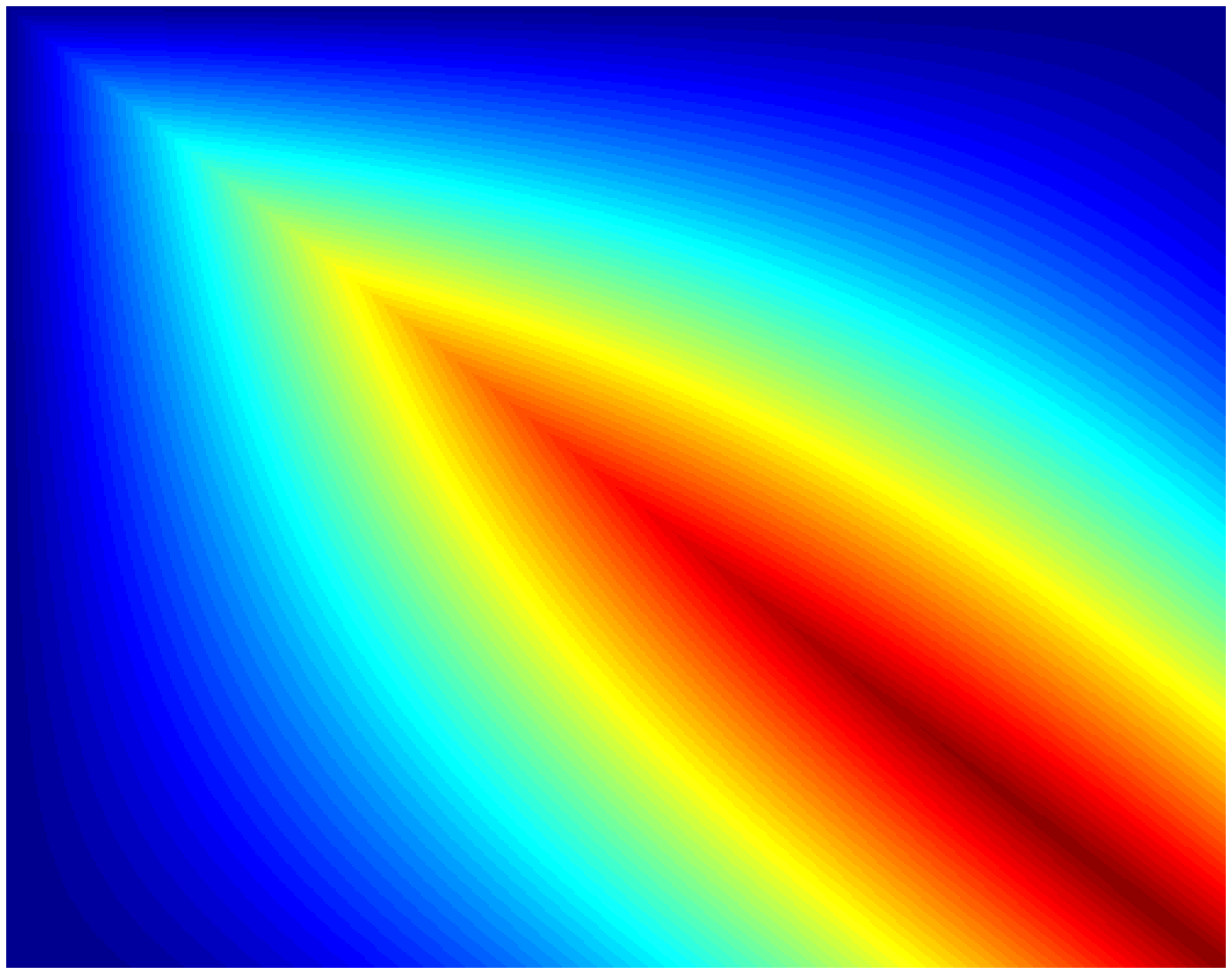}\quad
\includegraphics[height=4.5cm,width=4.5cm]{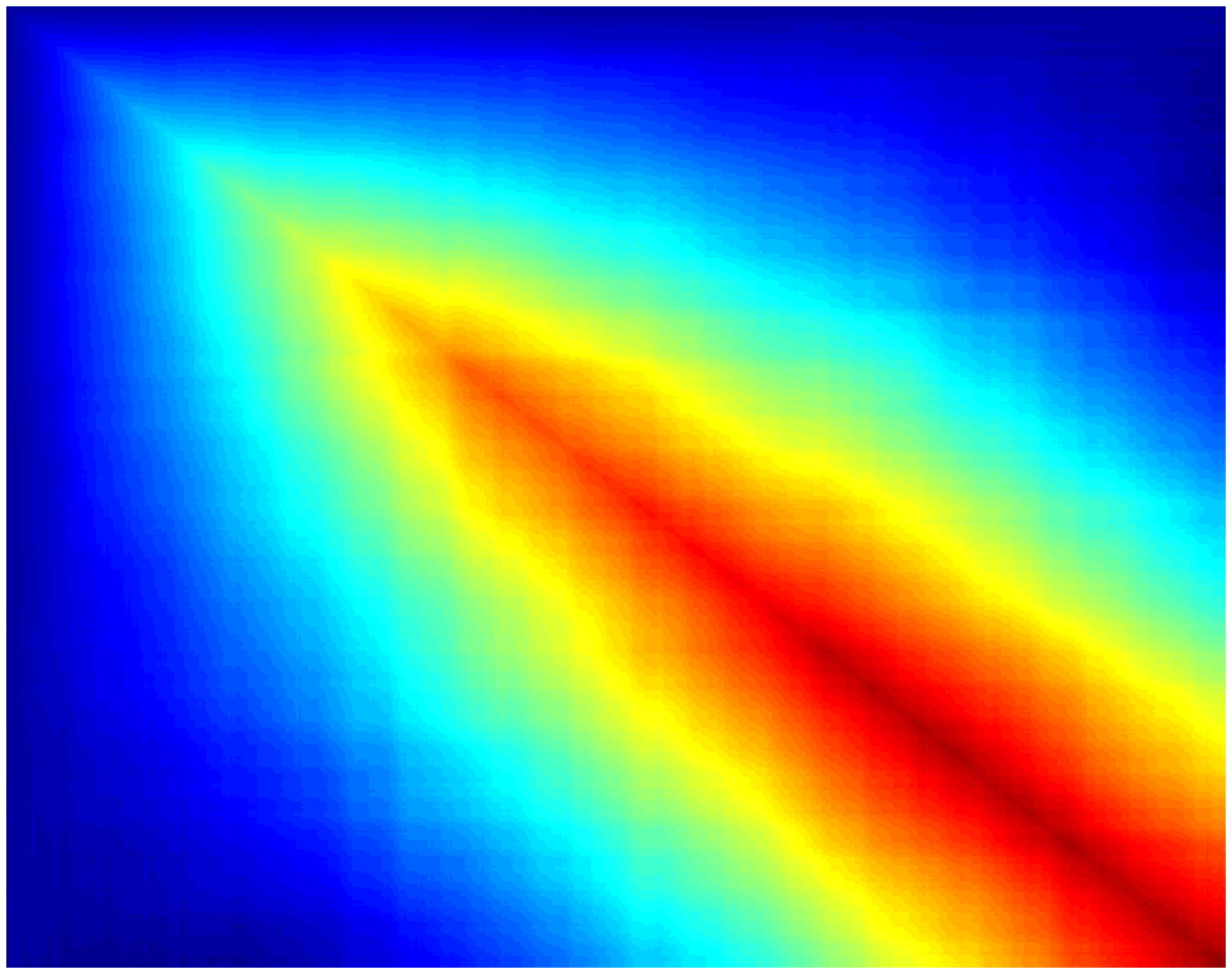}\\[.25cm]
\end{center}
{\bf Fig. 5:} Sample covariance matrix (left hand), true covariance matrix (middle), and spectral estimate
(right hand) of multivariate $t$-distributed realizations ($n=1000,\,d=500,\,\nu=4$).\\[.25cm]

\section{Random Matrix Theory}\label{RMT}

RMT is concerned with the distribution of the eigenvalues of high-dimensional randomly gene\-rated matrices. A
random matrix is simply a matrix of random variables. We will consider only symmetric random matrices. Thus the
corresponding eigenvalues are always real. The empirical distribution function of eigenvalues is defined as
follows.

\begin{definition}[Empirical distribution function of eigenvalues]
Let $\widehat{\Sigma}$ be a $d\times d$ symmetric random matrix with eigenvalues
$\widehat{\lambda}_{1},\widehat{\lambda}_{2},\ldots ,\widehat{\lambda}_{d}\,$. Then the function
\begin{equation*}
\lambda \longmapsto \widehat{W}_{d}\left( \lambda \right) :=\frac{1}{d}\cdot
\sum_{i=1}^{d}1\!\!1_{\widehat{\lambda}_{i}\leq \,\lambda }
\end{equation*}
is called the `empirical distribution function of the eigenvalues' of $\,\widehat{\Sigma}$.
\end{definition}

Note that each eigenvalue of a random matrix in fact is random but per se not a random variable since there is
no single-valued mapping $\widehat{\Sigma}\mapsto\widehat{\lambda}_{i}$ $\left( i\in \left\{ 1,\ldots ,d\right\}
\right)$ but rather $\widehat{\Sigma}\mapsto\lambda (\widehat{\Sigma})$ where $\lambda (\widehat{\Sigma})$
denotes the set of all eigenvalues of $\widehat{\Sigma}$. This can be simply fixed by assuming that the
eigenvalues $\widehat{\lambda}_{1},\widehat{\lambda}_{2},\ldots ,\widehat{\lambda}_{d}$ are sorted either in an
increasing or decreasing order.

\begin{theorem}[Mar\v{c}enko and Pastur, 1967]\label{MP_law}
Let $U_{1}^{\left( d\right) },U_{2}^{\left( d\right) },\ldots ,U_{n}^{\left( d\right) }$ $\left( n=1,2,\ldots
\right)$ be sequences of independent random vectors uniformly distributed on the unit hypersphere
$\mathcal{S}^{d-1}$ and consider the random matrix
\begin{equation*}
\widehat{\Sigma}_{\mathrm{MP}}:=\frac{d}{n}\cdot\sum_{j=1}^{n}U_{j}^{\left( d\right) }U_{j}^{\left( d\right)
\mathrm{T}},
\end{equation*}%
where its empirical distribution function of the eigenvalues is denoted by $%
\widehat{W}_{d}\,$. Suppose that $n\rightarrow \infty $,$\ d\rightarrow \infty $, $n/d\rightarrow q<\infty $.
Then
\begin{equation*}
\widehat{W}_{d}\overset{\mathrm{p}}{\longrightarrow }F_{\mathrm{MP}}\left(\cdot\,;q\right),
\end{equation*}
at all points where $F_{\mathrm{MP}}$ is continuous. More precisely, $\lambda \mapsto F_{\mathrm{MP}}\left(
\lambda \,;q\right) =F_{\mathrm{MP}}^{\mathrm{Dir}}\left( \lambda \,;q\right)
+F_{\mathrm{MP}}^{\mathrm{Leb}}\left( \lambda \,;q\right) $ where the Dirac part is given by
\begin{equation*}
\lambda \longmapsto F_{\mathrm{MP}}^{\mathrm{Dir}}\left( \lambda \,;q\right) =\left\{
\begin{array}{lll}
1-q, &  & \lambda \geq 0,\,0\leq q<1, \\
\rule{0cm}{0.5cm}0, &  & \text{else},%
\end{array}%
\right.
\end{equation*}%
and the Lebesgue part $\lambda \mapsto F_{\mathrm{MP}}^{\mathrm{Leb}}\left(
\lambda \,;q\right) =\int_{-\infty }^{\lambda }f_{\mathrm{MP}}^{\mathrm{Leb}%
}\left( x\,;q\right) dx$ is determined by the density function%
\begin{equation*}
\lambda \longmapsto f_{\mathrm{MP}}^{\mathrm{Leb}}\left( \lambda \,;q\right) =\left\{
\begin{array}{lll}
\frac{q}{2\pi}\cdot \frac{\sqrt{\left( \lambda _{\max }-\lambda \right) \left( \lambda -\lambda _{\min }\right)
}}{\lambda }, &  & \lambda _{\min }< \lambda < \lambda _{\max }, \\
\rule{0cm}{0.5cm}0, &  & \text{else},%
\end{array}%
\right.
\end{equation*}%
where%
\begin{equation*}
\lambda _{\min ,\max }:=\left( 1\pm \frac{1}{\sqrt{q}}\right) ^{2}.
\end{equation*}
\end{theorem}

\begin{proof}
Mar\v{c}enko and Pastur, 1967.\hfill \medskip
\end{proof}

In the following $\widehat{\Sigma}_{\mathrm{MP}}$ will be called `Mar\v{c}enko-Pastur operator'. The next
corollary states that the Mar\v{c}enko-Pastur law $F_{\mathrm{MP}}$ holds not only for the empirical
distribution function of eigenvalues of the Mar\v{c}enko-Pastur operator but also for that obtained by the
sample covariance matrix if the data are standard normally distributed and independent.

\begin{corollary}
Let $X,X_{1},X_{2},\ldots ,X_{n}$ $\left( n=1,2,\ldots \right)$ be sequences of independent and standard normally
distributed random vectors with uncorrelated components. Then the empirical distribution function of the eigenvalues of
\begin{equation*}
\frac{1}{n}\cdot\sum_{j=1}^{n}X_{j}X_{j}^{\mathrm{T}}
\end{equation*}
converges in probability to the Mar\v{c}enko-Pastur law stated in Theorem \ref{MP_law}.
\end{corollary}

\begin{proof}
Due to the strong law of large numbers $\chi _{d}^{2}/d\overset{\mathrm{a.s.}}{\rightarrow }1$
$(d\rightarrow\infty)$ and thus
\begin{equation*}
\widehat{\Sigma}_{\mathrm{MP}} \sim \frac{d}{n}\cdot \sum_{j=1}^{n}\frac{\chi _{d,j}^{2}}{d}\cdot U_{j}^{\left(
d\right) }U_{j}^{\left( d\right) \mathrm{T}} \overset{\mathrm{d}}{=}
\frac{1}{n}\cdot\sum_{j=1}^{n}X_{j}X_{j}^{\mathrm{T}}.
\end{equation*}
\rule{.5cm}{0cm}\hfill\medskip
\end{proof}

Moreover, the Mar\v{c}enko-Pastur law holds even if $X$ is an arbitrary random vector with standardized i.i.d.
components provided the second moment is finite (Yin, 1986). More precisely, consider the random vector $X$ with
$E(X)=\mu$ and $Var(X)=\sigma^2 I_{d}$ where the components of $X$ are supposed to be stochastically
independent. Then the Mar\v{c}enko-Pastur law can be applied on the empirical distribution function of the
eigenvalues of
\begin{equation*}
\frac{1}{n}\cdot\sum_{j=1}^{n}\left(\frac{X_{j}-\widehat{\mu}}{\widehat{\sigma}}\right)
\left(\frac{X_{j}-\widehat{\mu}}{\widehat{\sigma}}\right)^{\mathrm{T}}= \widehat{\Sigma}/\widehat{\sigma}^2,
\end{equation*}
where $\widehat{\Sigma}$ denotes the sample covariance matrix and
\begin{equation*}
\widehat{\sigma}^2:=\frac{\tr(\widehat{\Sigma})}{d}=\frac{1}{d}\cdot
\sum_{i=1}^{d}\widehat{\lambda}_{i}=:\overline{\lambda}.
\end{equation*}

Hence, the Mar\v{c}enko-Pastur law can be applied virtually ever on the empirical distribution function of
$\widehat{\lambda}_{1}/\overline{\lambda},...,\widehat{\lambda}_{d}/\overline{\lambda}$ where the estimated
eigenvalues are given by the sample covariance matrix provided the sample elements, i.e. the realized random
vectors consist of stochastically independent components. But within the class of elliptical distributions this
holds only for uncorrelated normally distributed data. Hence linear independence and stochastical independence
are not equivalent for genera\-lized elliptically distributed data. This is because even if there is no linear dependence between the components of an elliptically distributed random vector another sort of nonlinear dependence caused by the gene\-rating variate $\mathcal{R}$ remains, generally.

For instance, consider the unit random vector
$U^{(2)}=(U_{1},U_{2})$. Then
\begin{equation*}
U_{2}\overset{\mathrm{a.s.}}{=}\pm \sqrt{1-U_{1}^{2}},
\end{equation*}%
i.e. $U_{2}$ depends strongly on $U_{1}$ though indeed the elements of $U^{(2)}$ are uncorrelated.

Tail dependent random variables cannot be stochastically independent. Especially, if the random components of an elliptically distributed random vector are heavy tailed, i.e. if the generating variate is regularly varying then they possess the property of tail dependence (Schmidt, 2002). In that case the eigenspectrum generated by the sample covariance matrix may lead to erroneous implications.

For instance, consider a sample (with sample size $n=1000$) of $500$-dimensional random vectors where each vector element is standardized $t$-distributed with $\nu=5$ degrees of freedom and stochastically independent of each other. Here the eigenspectrum obtained by the sample covariance matrix indeed is consistent with the Mar\v{c}enko-Pastur law (upper left part of Figure 6). But if the data stem from a multivariate $t$-distribution possessing the same parameters and each vector component is uncorrelated then the eigenspectrum obtained by the sample covariance matrix does not correspond to the Mar\v{c}enko-Pastur law (upper right part of Figure 6). Actually, there are $24$ eigenva\-lues exceeding the Mar\v{c}enko-Pastur upper bound $\lambda _{\max}=(1+1/\sqrt{2}\,)^{2}=2.91$ and the largest eigenvalue corresponds to $10.33$. But fortunately
the eigenspectra obtained by the spectral estimator are consistent with the Mar\v{c}enko-Pastur law as
indicated by the lower part of Figure 6.

\begin{center}
\includegraphics[scale=.34]{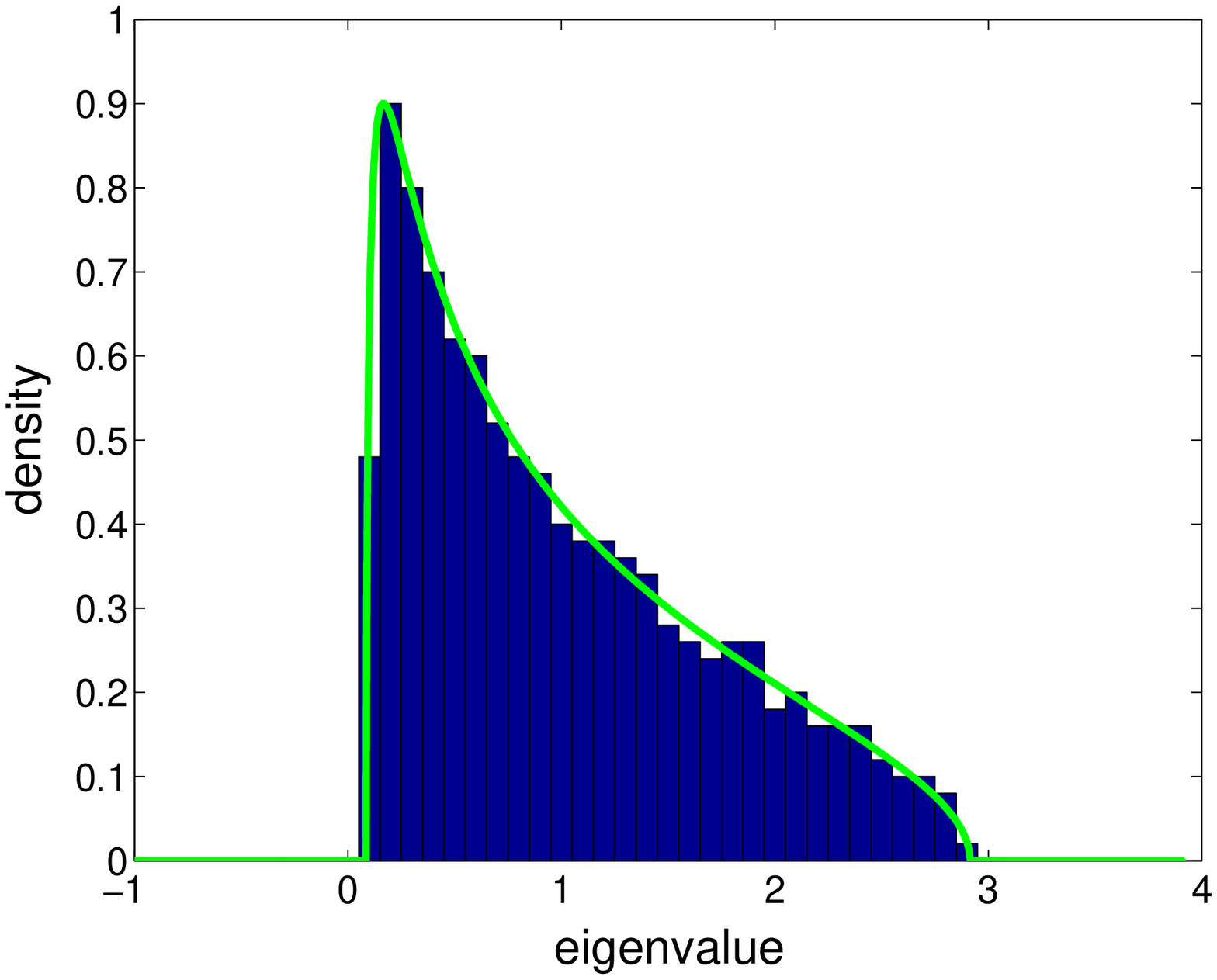}
\includegraphics[scale=.34]{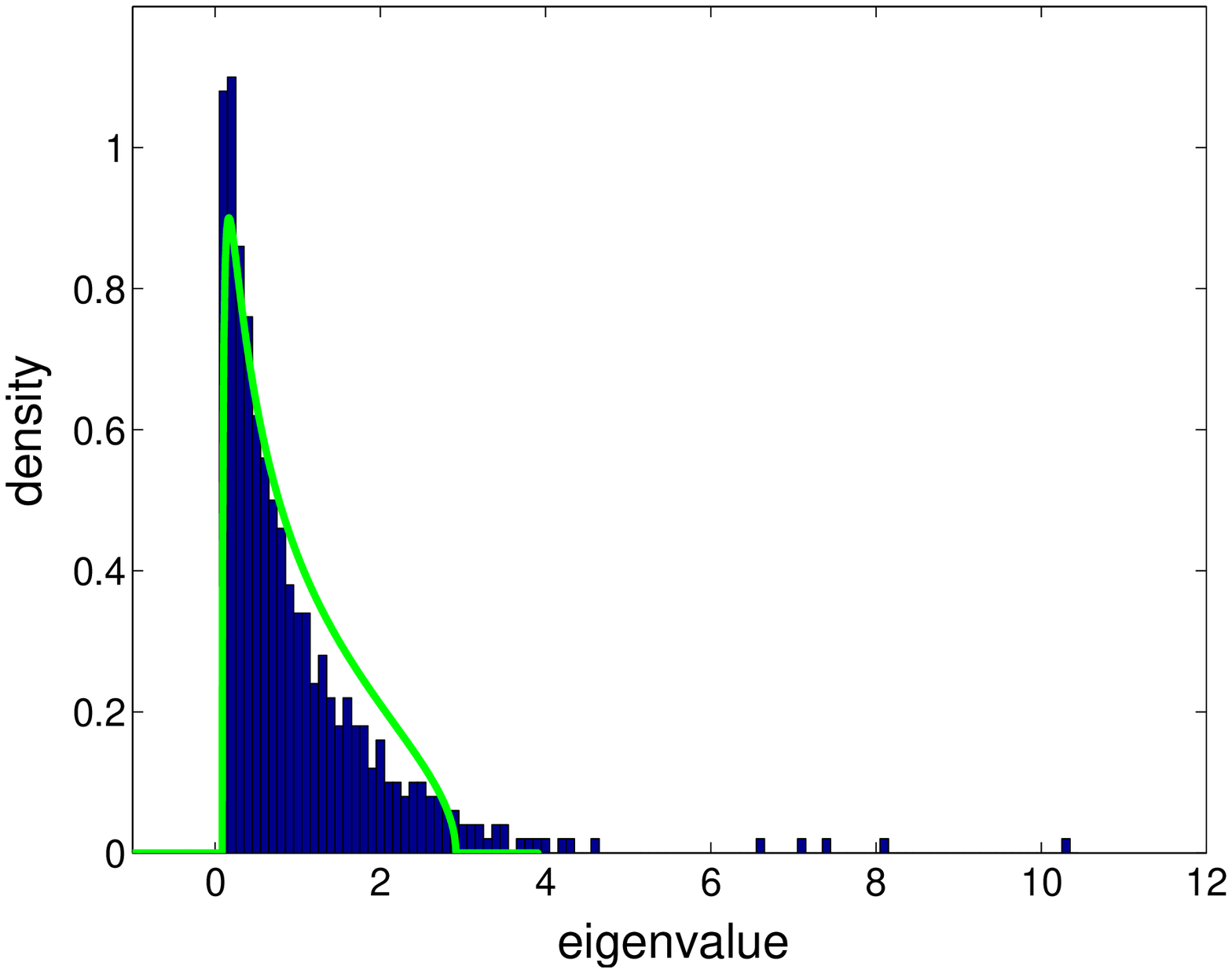}\\[.25cm]
\includegraphics[scale=.34]{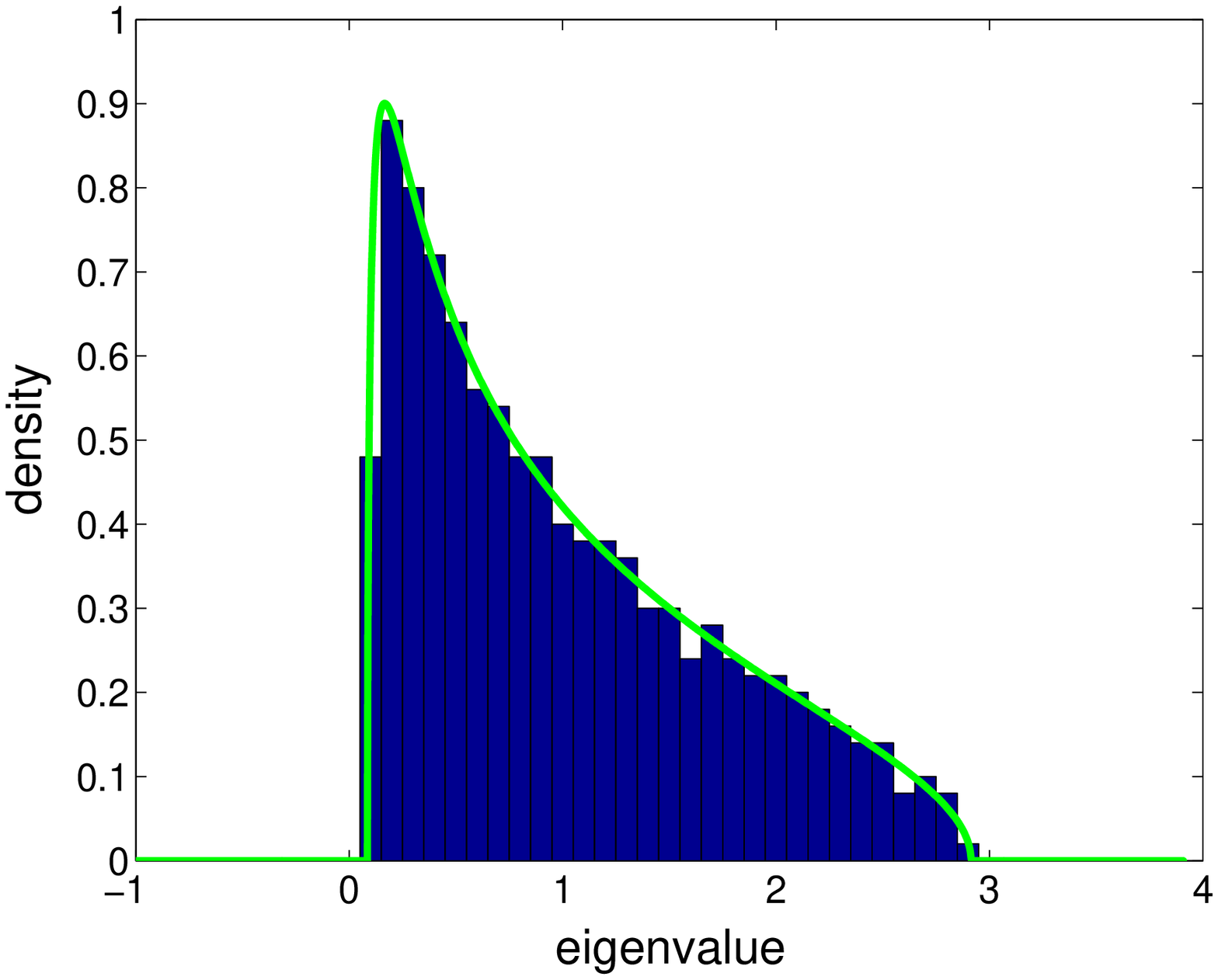}
\includegraphics[scale=.34]{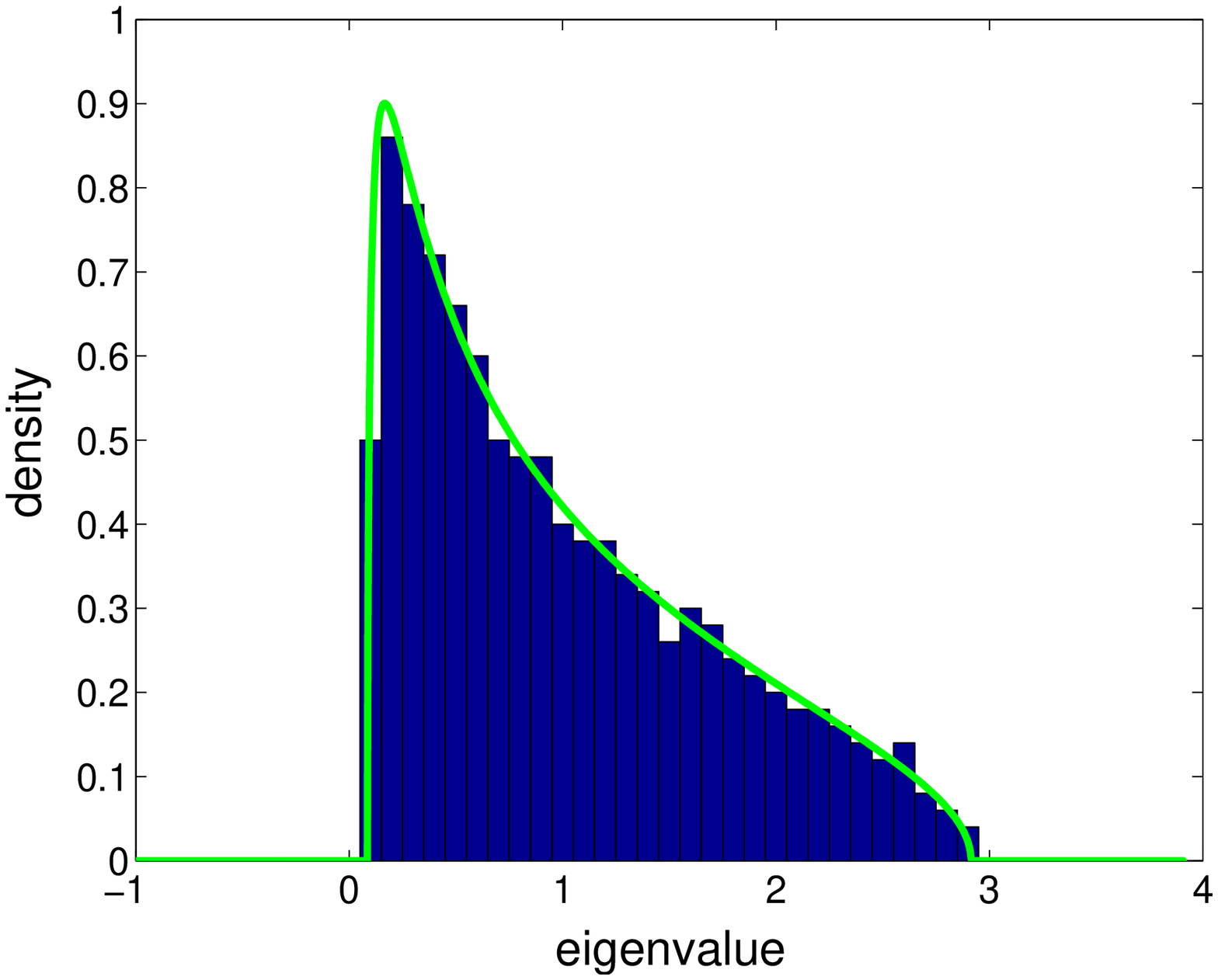}\\[.25cm]
\end{center}
{\bf Fig. 6:} Eigenspectra of univariate (left part) and multivariate (right part) uncorrelated $t$-distributed
data ($n=1000,\,d=500,\,\nu=5$) obtained by the sample covariance matrix (upper part) and by the spectral
estimator (lower part).\\[.25cm]

Tyler (1987) shows that the spectral estimator converges strongly to the true dispersion matrix $\Sigma $. That means %
\begin{equation*}
\frac{s_{j}s_{j}^{\mathrm{T}}}{s_{j}^{\mathrm{T}}%
\widehat{\Sigma }^{-1}s_{j}}\longrightarrow \frac{%
s_{j}s_{j}^{\mathrm{T}}}{s_{j}^{\mathrm{T}}\Sigma ^{-1}s_{j}},\qquad n\longrightarrow \infty ,\ d\text{ const.,}
\end{equation*}%
for $j=1,2,\ldots$ and $P$-almost all realizations. Consequently, if $\Sigma =I_{d}$ (up to a scaling constant) then%
\begin{equation*}
\frac{s_{j}s_{j}^{\mathrm{T}}}{s_{j}^{\mathrm{T}}%
\widehat{\Sigma }^{-1}s_{j}}\longrightarrow s_{j}s_{j}^{\mathrm{T}} \equiv u_{j}^{\left(d\right)}u_{j}^{\left(d\right)\mathrm{T}},
\end{equation*}%
as $n\rightarrow\infty$ and $d$ constant. Hence the spectral estimator and the Mar\v{c}enko-Pastur operator are
asymptotically equivalent provided $\Sigma =\sigma^{2}I_{d}$. The authors believe that the strong
convergence holds even for $n\rightarrow \infty $, $d\rightarrow \infty $, $n/d\rightarrow q>1$ for $P$-almost
all realizations where the spectral estimate exists. The proof of this conjecture is due to a forthcoming work.
Note that for $q\leq 1$ the spectral estimate does not exist at all. Further, Tyler (1987) shows that the
spectral estimate exists (a.s.) if $n>d\left(d-1\right)$, i.e. $q>d-1$. Indeed, this is a sufficient condition
for the existency of the spectral estimator. But in practice the spectral estimator seems to exist in most cases
when $n$ is already slightly larger than $d$.

We conclude that testing high-dimensional data for the null hypothesis $\Sigma =\sigma^{2}I_{d}$ by means of the sample covariance matrix may lead to wrong conclusions provided the data are generalized elliptically distributed. In contrast, the spectral estimator seems to be a robust alternative for applying the results of RMT in the context of generalized elliptical distributions.

\section{Financial Applications}

\subsection{Portfolio Risk Minimization}

In this section it is supposed that $n/d\rightarrow \infty$, i.e. from the viewpoint of RMT we
study low-dimensional problems. Let $R=(R_{1},R_{2},...,R_{d})$ be an elliptically distributed random vector of
short-term (e.g. daily) log-returns. If the fourth order cross moments of the log-returns are
finite then the elements of the sample covariance matrix are multivariate normally distributed, asymptotically.
The asymptotic covariance of each element is given by (see, e.g., Praag and Wesselman, 1989)
\begin{equation*}
\mathrm{ACov}\left(\hat{\sigma}_{ij},\hat{\sigma}_{kl}\right) =\left( 1+\kappa \right) \cdot \left( \sigma
_{ik}\sigma _{jl}+\sigma _{il}\sigma _{jk}\right) +\kappa\cdot\sigma _{ij}\sigma _{kl},
\end{equation*}
where $\Sigma=[\sigma_{ij}]$ denotes the true covariance matrix of $R$ and
\begin{equation*}
\kappa :=\frac{1}{3}\cdot \frac{E\left( R_{i}^{4}\right) }{E^{2}\!\left( R_{i}^{2}\right) }-1
\end{equation*}
is called the `kurtosis parameter'. Note that the kurtosis parameter does not depend on $i\in\{1,...,d\}$. It
is well-known that in the case of normality $\kappa =0$. A distribution with positive (or even infinite)
$\kappa $ is called `leptokurtic'. Particularly, regularly varying distributions are leptokurtic.

It is well-known that the portfolio which minimizes the portfolio return variance (the so called `global minimum
variance portfolio') is given by the vector of portfolio weights
\begin{equation*}\label{GMVP}
w := \frac{\Sigma ^{-1}\text{$\underline{1}$}}{\text{$\underline{1}$}^{\mathrm{T}}\Sigma
^{-1}\text{$\underline{1}$}}.
\end{equation*}

Now, suppose for the sake of simplicity that $R$ is spherically distributed, i.e. that $\mu = 0$ and $\Sigma$ is proportional to the identity matrix. Since the weights of the global minimum variance portfolio do not depend on the scale of $\Sigma$ we may assume $\Sigma = I_{d}$ w.l.o.g. Then the asymptotic covariances of the sample covariance matrix elements are simply given by
\begin{equation*}
\mathrm{ACov}\left(\hat{\sigma}_{ij},\hat{\sigma}_{kl}\right) =\left\{
\begin{array}{rcl}
2+3\kappa , &  & i=j=k=l, \\ \rule{0cm}{0.5cm}\kappa , &  & i=j,\, k=l,\, i\neq k, \\ \rule{0cm}{0.5cm}1+\kappa
, &  & i=k,\, j=l,\, i\neq j, \\ \rule{0in}{0.5cm}0, &  & \text{else}.
\end{array}\right.
\end{equation*}

For instance suppose that the random vector $R$ is multivariate $t$-distributed with $\nu>4$ de\-grees of
freedom. Then the kurtosis parameter corresponds to $\kappa =2/(\nu -4)$ (see, e.g., Frahm, 2004, p. 91).
Hence, the smaller $\nu$ the larger the asymptotic variances and covariances and these quantities tend to
infinity for $\nu \searrow 4$. Further, if $\nu\leq 4$ the sample covariance matrix even is no longer
multivariate nor\-mally distributed, asymptotically.

In contrast, the asymptotic covariance of each element of the spectral estimator (Frahm, 2004, p.
76) is given by
\begin{equation*}
\mathrm{ACov}\left(\hat{\sigma}_{\mathrm{S},ij},\hat{\sigma}_{\mathrm{S},kl}\right) =\left\{
\begin{array}{rcl}
4\cdot\frac{d+2}{d} , &  & i=j=k=l, \\ \rule{0cm}{0.5cm}2\cdot\frac{d+2}{d} , &  & i=j,\, k=l,\, i\neq k, \\
\rule{0cm}{0.5cm}\frac{d+2}{d} , &  & i=k,\, j=l,\, i\neq j, \\ \rule{0in}{0.5cm}0, &  & \text{else}.
\end{array}\right.
\end{equation*}

Note that the same holds even if $R$ is not $t$-distributed but only generalized elliptically distributed since
$\widehat{\Sigma}_{\mathrm{S}}$ does not depend on the generating variate of $R$. Particularly, the spectral
estimator is not disturbed by the tail index of $R$.

Now one may ask when the sample covariance matrix is dominated (in a component-wise manner) by the spectral
estimator provided the data are multivariate $t$-distributed. Regarding the main diagonal entries of the
covariance matrix estimate this is given by
\begin{equation*}
4\cdot \frac{d+2}{d}<2\cdot \frac{\nu -1}{\nu -4},
\end{equation*}
i.e. if $\nu <4 + 3d/(d+4)$ the variance of the spectral estimator's main diagonal elements is smaller than
the variance of the corresponding main diagonal elements of the sample covariance matrix, asymptotically. Concerning its off diagonal entries we obtain
\begin{equation*}
\frac{d+2}{d}<\frac{\nu -2}{\nu -4},
\end{equation*}
i.e. $\nu < 4+d$. It is worth to note that several empirical studies indicate that the tail indices of daily log-returns generally lie between $4$ and $7$ (see, e.g., Embrechts, Frey, and McNeil, 2004, p. 81 and Junker and May, 2002).

In the following the daily log-returns from 1980-01-02 to 2003-10-06 of 285 S\&P 500 stocks are analyzed for studying the robustness of the spectral estimator vs. the sample covariance matrix. The considered stocks belong to the `survivors' of the S\&P 500 composite at the last quarter of 2003. The sample size corresponds to $n=6000$. The total sample period is partitioned into $10$ sub-periods each containing $600$ daily log-returns. Further, each sub-period is divided into `even' and `odd' days, i.e. there is a sub-sample containing the 1st, 3rd, \ldots, 599th log-returns and another sub-sample with the 2nd, 4th, \ldots, 600th log-returns. Hence each sub-sample contains $300$ daily log-returns of $285$ stocks. Both the sample covariance matrix and the spectral estimator are used for estimating the relative eigenspectrum of the true covariance matrix, i.e. $\lambda_{1}/\sum_{i=1}^{d}\lambda_{i},\ldots ,\lambda_{d}/\sum_{i=1}^{d}\lambda_{i}$ for each even and odd sub-sample, separately. If the covariance matrix estimator is robust against outliers then the estimated eigenspectra of each sub-sample should be similar since even if the true eigenspectrum changes dynamically over time this must affect both the even and the odd days, equally. The eigenspectrum obtained in the even sub-sample can be compared with the eigenspectrum given by the odd sub-sample simply by the differences of the ordered (relative) eigenvalues.

\begin{center}
\includegraphics[scale=.21]{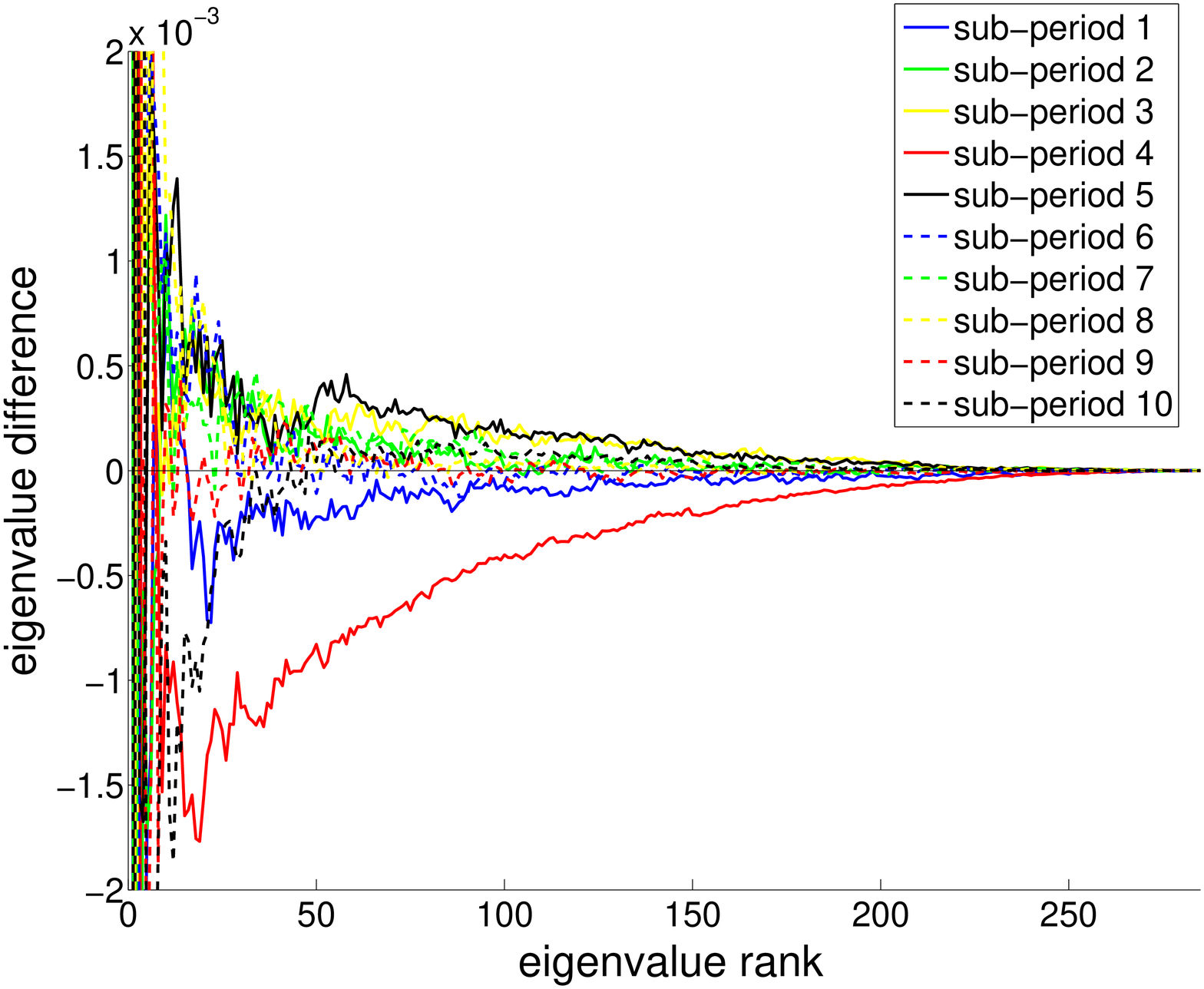}
\includegraphics[scale=.21]{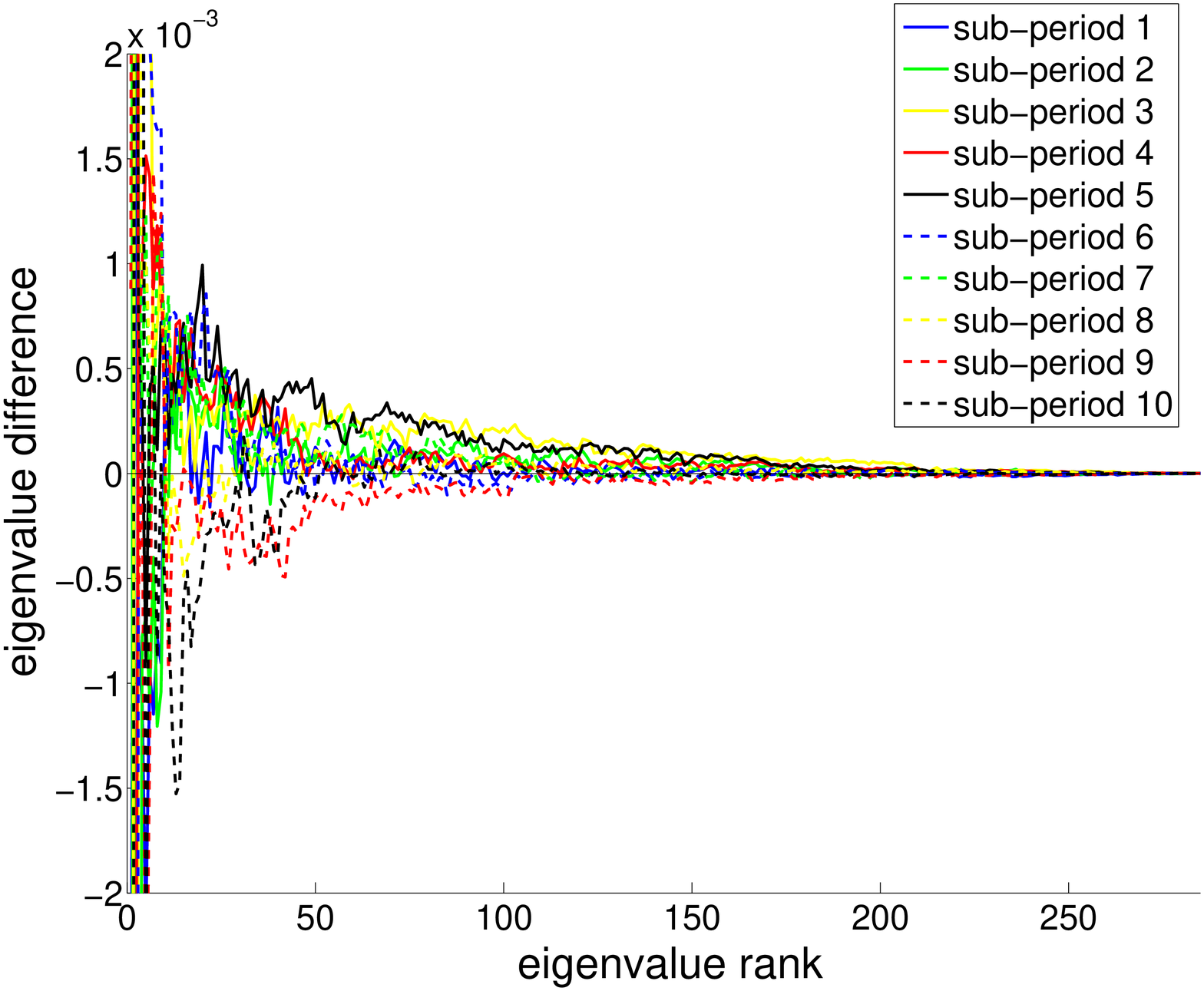}\\[.25cm]
\end{center}
{\bf Fig. 7:} Eigenvalue differences for each ordered eigenvalue given by the sample covariance matrix (left hand) and by the spectral estimate (right hand).\\[.25cm]

On the left hand of Figure 7 we see the eigenvalue differences for each $10$ sub-periods caused by the sample covariance matrix. Similarly, the right hand of Figure 7 shows the eigenvalue differences given by the spectral estimate. Figure 7 indicates that the spectral estimator leads to more robust estimates of the eigenspectra of financial data. But note that - concerning the overall eigenspectrum - the sample covariance matrix performs well up to the 4th sub-period. This is the period which contains the famous October Crash of $1987$. In contrast, the spectral estimator is not affected by extreme values.

\begin{center}
\includegraphics[scale=.21]{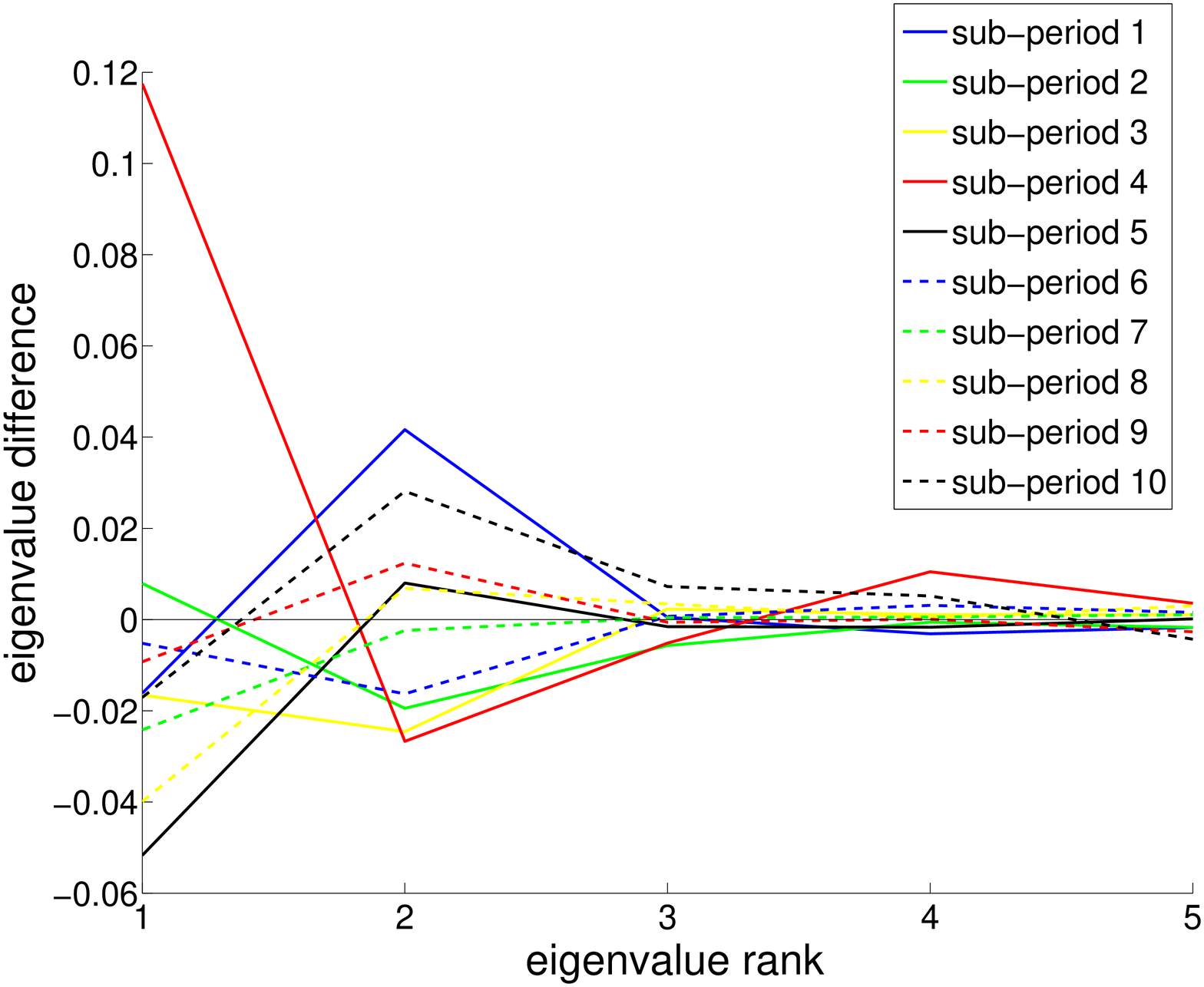}
\includegraphics[scale=.21]{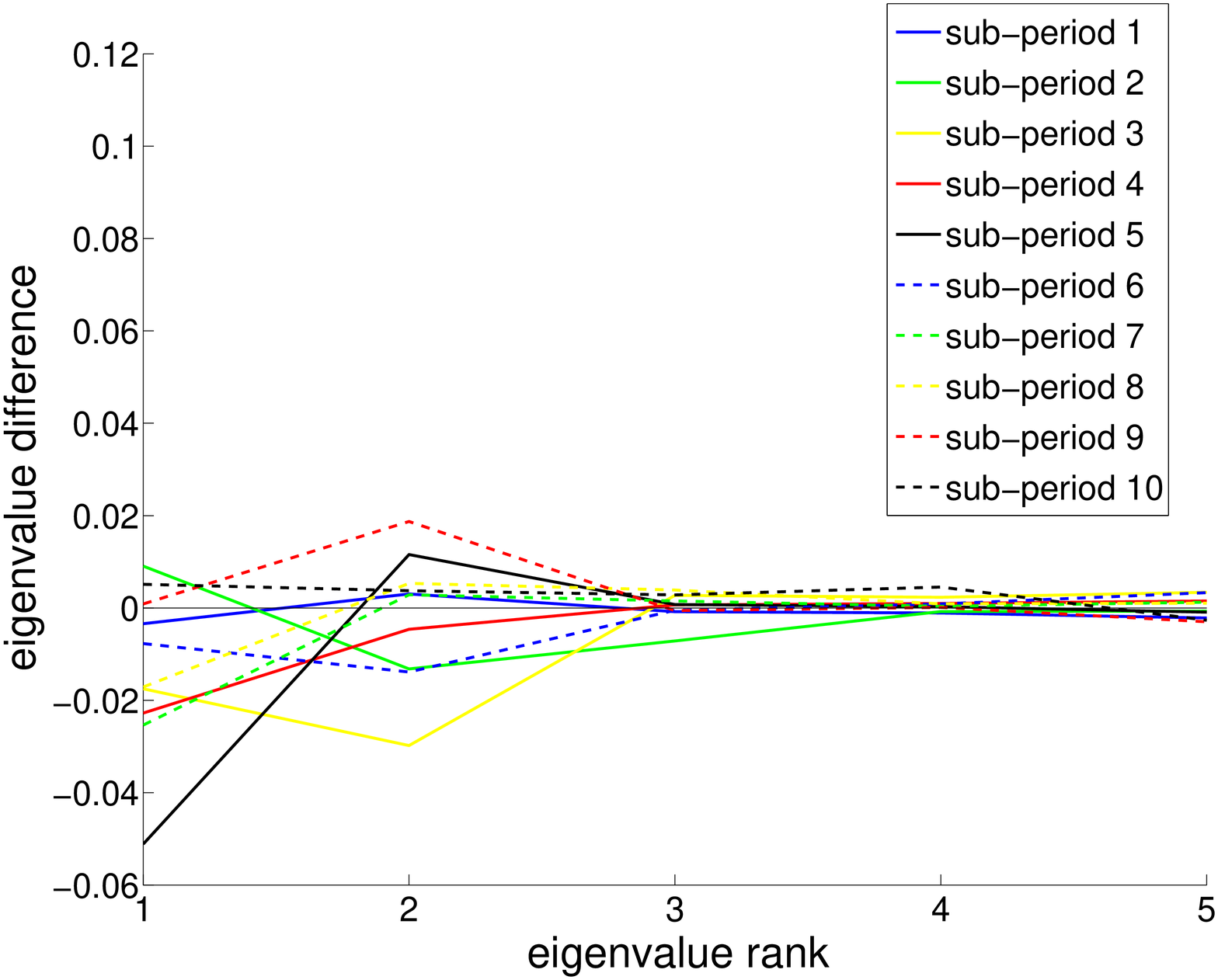}\\[.25cm]
\end{center}
{\bf Fig. 8:} Eigenvalue differences for the largest $5$ eigenvalues given by the sample covariance matrix (left hand) and by the spectral estimate (right hand).\\[.25cm]

Figure 8 focuses on the differences of the $5$ largest eigenvalues. It shows that the sample covariance matrix particularly fails for estimating the largest eigenvalue. Once again this phenomenon is caused by the Black Monday which belongs to the even sub-sample of the 4th sub-period. Note that the largest eigenvalue of the even sub-sample exceeds the largest eigenvalue of the odd sub-sample by almost $12$ percentage points. We conclude that although the sample covariance matrix works quite good for the most time it is not appropriate for measuring the linear dependence structure of financial data. This is due to a few but extreme fluctuations on financial markets.

\subsection{Principal Components Analysis}

Now, consider a $d$-dimensional vector $R=(R_{1},...,R_{d})$ of long-term (e.g. yearly) i.i.d. log-returns. Due to the central limit theorem each vector component of $R$ is approximately normal distributed provided the covariance matrix of the short-term (e.g. daily) log-returns exists and is finite. Since the sum of i.i.d. elliptical random vectors is always elliptically distributed, too (see, e.g., Hult and Lindskog, 2002) one may take for granted that the vector components of $R$ are jointly normally distributed, approximately. But this is not true if the number of dimensions $d$ is large relative to the sample size $n$.

For instance, consider a $d$-dimensional random vector $X$ which is multivariate $t$-distributed with $\nu>2$ degrees of freedom, location vector $\mu = 0$, and dispersion matrix $\Sigma = (\nu -2)/\nu\cdot I_{d}$. Due to the multivariate central limit theorem one could believe that
\begin{equation*}
Y := \frac{1}{\sqrt{n}}\cdot\sum_{j=1}^{n} X_{j}\overset{\cdot}{\sim}N_{d}\left( 0,I_{d}\right),
\end{equation*}
where $X_{1},\ldots,X_{n}$ are independent copies of $X$. But indeed $Y^{\text{T}}Y \overset{\cdot}{\sim}\chi_{d}^{2}$ holds only if $q:=n/d$ is large rather than $n$ being large (cf. Frahm, 2004, Section 6.2). Thus the quantity $q$ can be interpreted as `effective sample size'.

In the following it is assumed that $R$ is elliptically distributed with location vector $\mu$ and dispersion matrix $\Sigma$. Let $\Sigma = \mathcal{O}\mathcal{D}\mathcal{O}^{\text{T}}$ be a spectral decomposition of $\Sigma$. Then
\begin{equation*}
R\overset{\mathrm{d}}{=}\mu +\mathcal{O}\sqrt{\mathcal{D}}\,Y,
\end{equation*}
where $Y$ spherically distributed with $\Sigma = I_{d}$.

We assume that the elements of $\mathcal{D}$, i.e. the eigenvalues of $\Sigma$ are given in a descending order
and that the first $k$ eigenvalues are large whereas the residual ones are small. The elements of $Y$ are called `principal components' of $R$. Since $\mathcal{O}$ is orthonormal the distribution of $\sqrt{\mathcal{D}}\,Y$ remains up to a rotation in $\R^{d}$. The direction of each principal component is given by the corresponding column of $\mathcal{O}$.

Hence the first $k$ eigenvalues correspond to the variances (up to a scaling constant) of the `driving risk factors' contained in the first part of $Y$, i.e. $\left( Y_{1},\ldots,Y_{k}\right)$. For the purpose of dimension reduction $k$ shall not be too large. Because the $d-k$ residual risk factors contained in $\left( Y_{k+1},\ldots
,Y_{d}\right) $ are supposed to have (relatively) small variances they can be interpreted as the components of the
idiosyncratic risks of each firm, i.e.
\begin{equation*}
\varepsilon _{i}:=\sum_{j=k+1}^{d}\sqrt{\lambda_{j}}\,\mathcal{O}_{ij}Y_{j},\qquad i=1,\ldots ,d,
\end{equation*}
where $\lambda_{j}:=\mathcal{D}_{jj}$.

Thus we obtain the following principal components model for long-term log-returns,
\begin{equation*}
R_{i}\overset{\mathrm{d}}{=}\mu_{i}+\beta _{i1}Y_{1}+\ldots +\beta _{ik}Y_{k}+\varepsilon _{i},\qquad
i=1,\ldots ,d,
\end{equation*}
where the driving risk factors $Y_{1},...,Y_{k}$ are uncorrelated. Further, each noise term
$\varepsilon_{i}$ $(i=1,...,d)$ is uncorrelated to $Y_{1},...,Y_{k}$, too. But note that $\varepsilon_{1},\ldots ,\varepsilon_{d}$ are correlated, generally. The `Betas' are given by $\beta_{ij} = \sqrt{\lambda_{j}}\,\mathcal{O}_{ij}$ for $i=1,\ldots , d$ and $j=1,\ldots ,k$.

The purpose of principal components analysis is to reduce the complexity caused by the number of dimensions.
This can be done successfully only if there is indeed a number of principal components accountable for the most
part of the distribution. Additionally, the covariance matrix estimator which is used for extracting the
principal components should be robust against outliers.

For example, let the daily log-returns be multivariate $t$-distributed with $\nu$ degrees of freedom and suppose
that $d=500$ and $n=1000$. Note that due to the central limit theorem the normality assumption concerning the long-term log-returns makes sense whenever $\nu >2$. The black lines in Figure 9 show the true proportion of the total variation for a set of $500$ eigenvalues. We see that the largest $20\%$ of the eigenvalues accounts for $%
80\%$ of the overall variance. This is known in economics as `80/20 rule' or `Pareto's principle'. The estimated
eigenvalue proportions obtained by the sample covariance matrix are represented by the red lines whereas the
corres\-ponding estimates based on the spectral estimator are given by the green lines. Each line is an average
over $100$ concentration curves drawn from samples of the corresponding multivariate $t$-distribution.

If the data have a small tail index as given by the lower right of Figure 9 then the sample covariance matrix
tends to underestimate the number of driving risk factors, essentially. This is similar to the phenomenon
observed in Figure 6 where the number of large eigenvalues is overestimated. In contrast, the concentration
curves obtained by the spectral estimator are robust against heavy tails. This holds even if the long-term log-returns are not asymptotically normal distributed.

\begin{center}
\includegraphics[scale=.34]{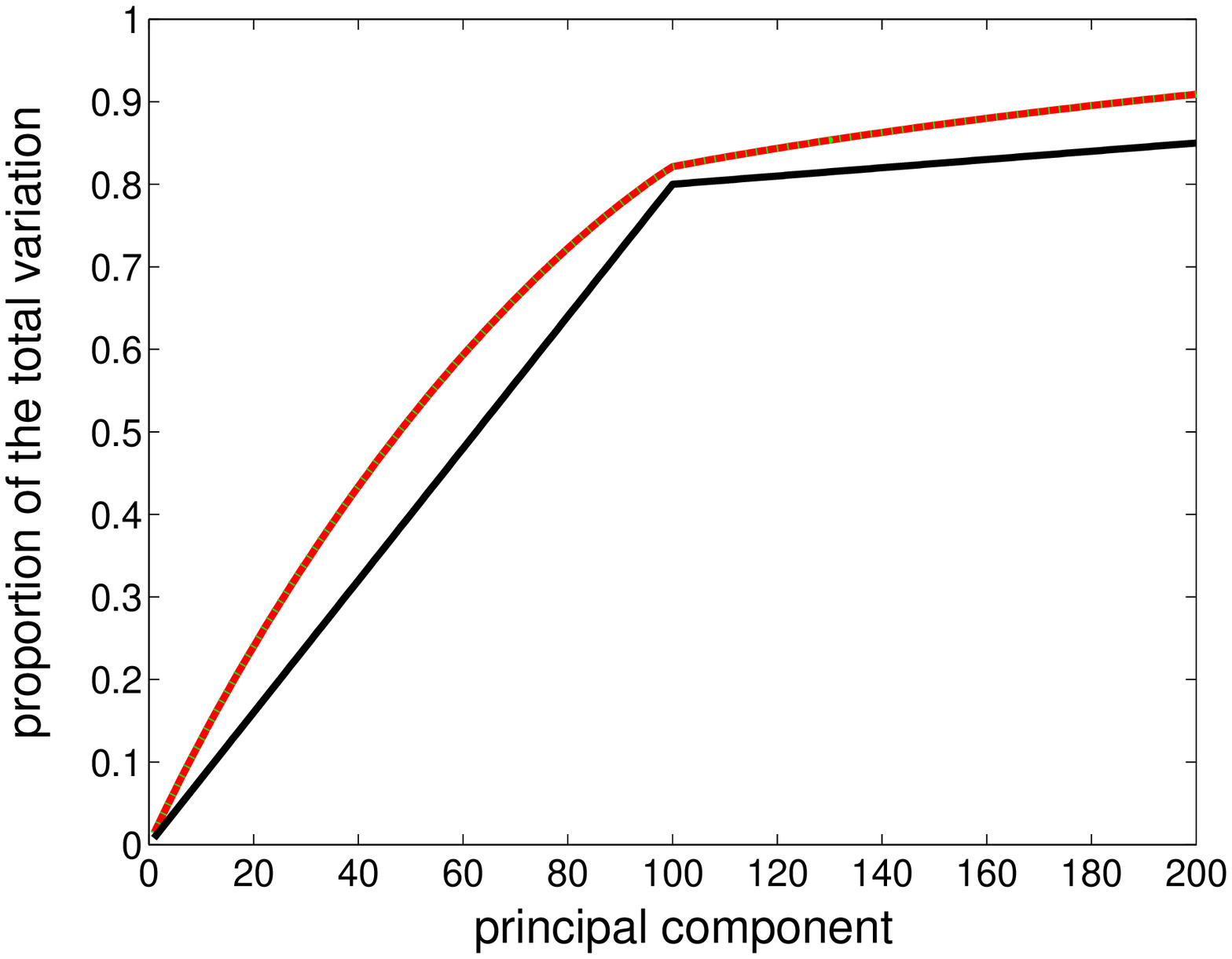}
\includegraphics[scale=.34]{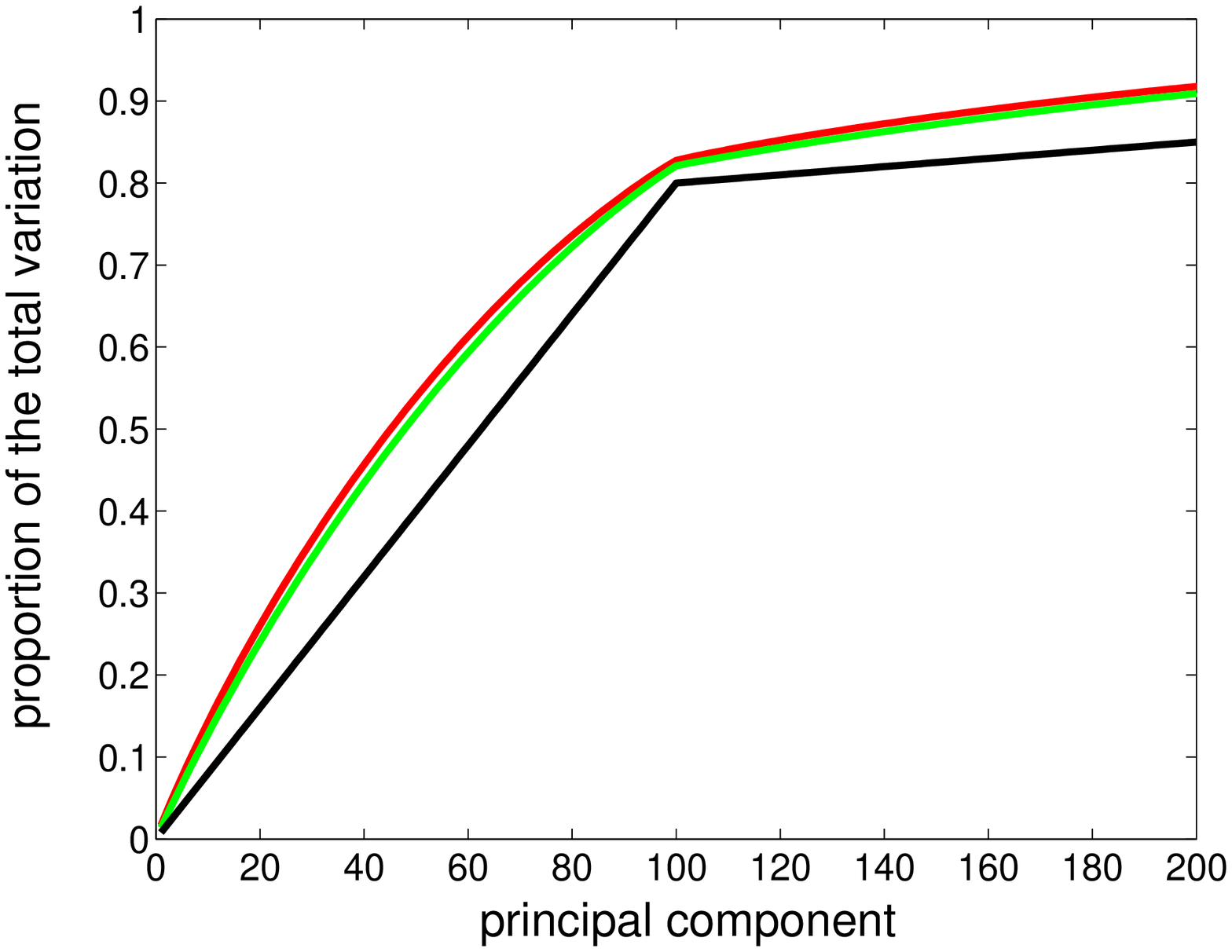}\\[.25cm]
\includegraphics[scale=.34]{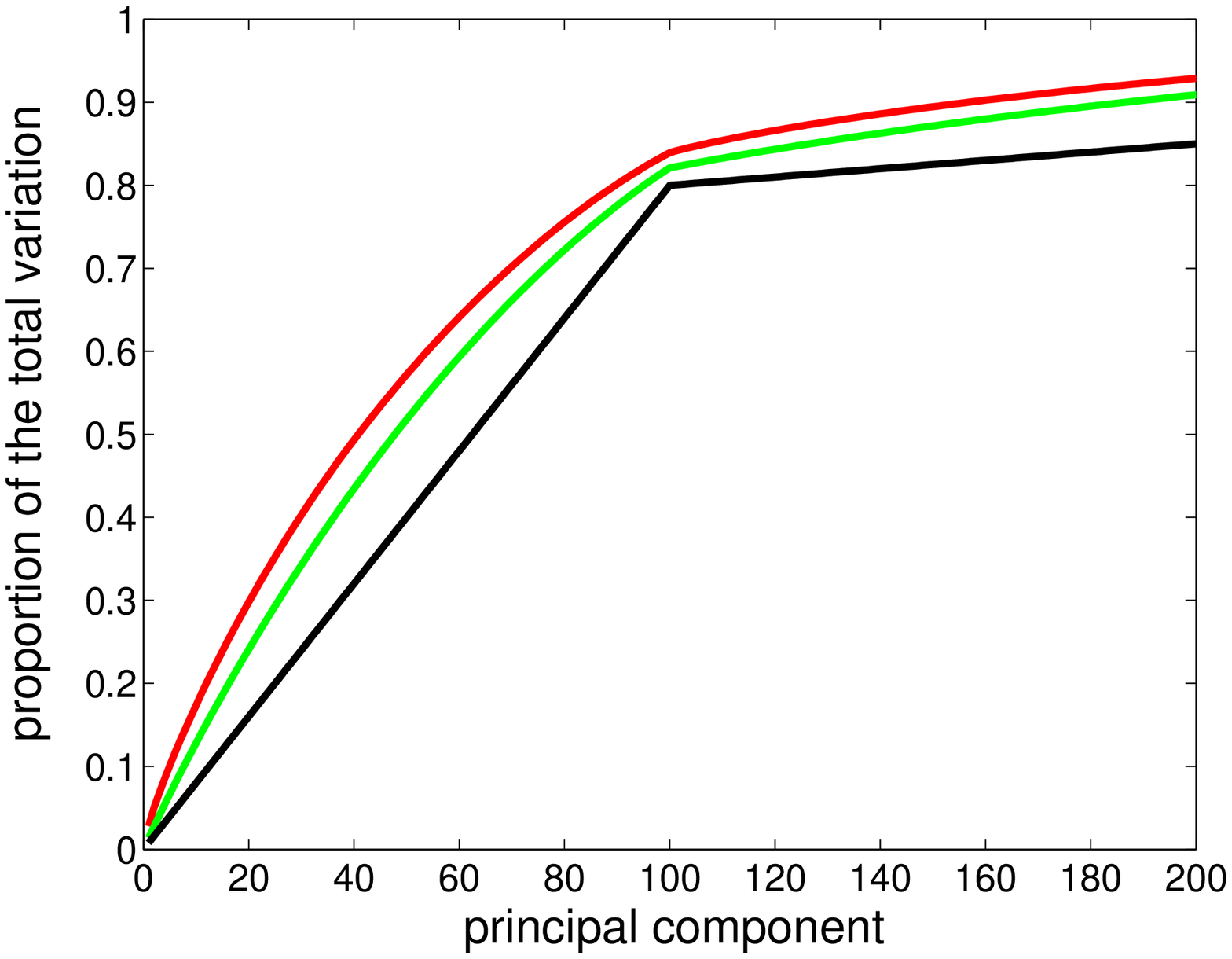}
\includegraphics[scale=.34]{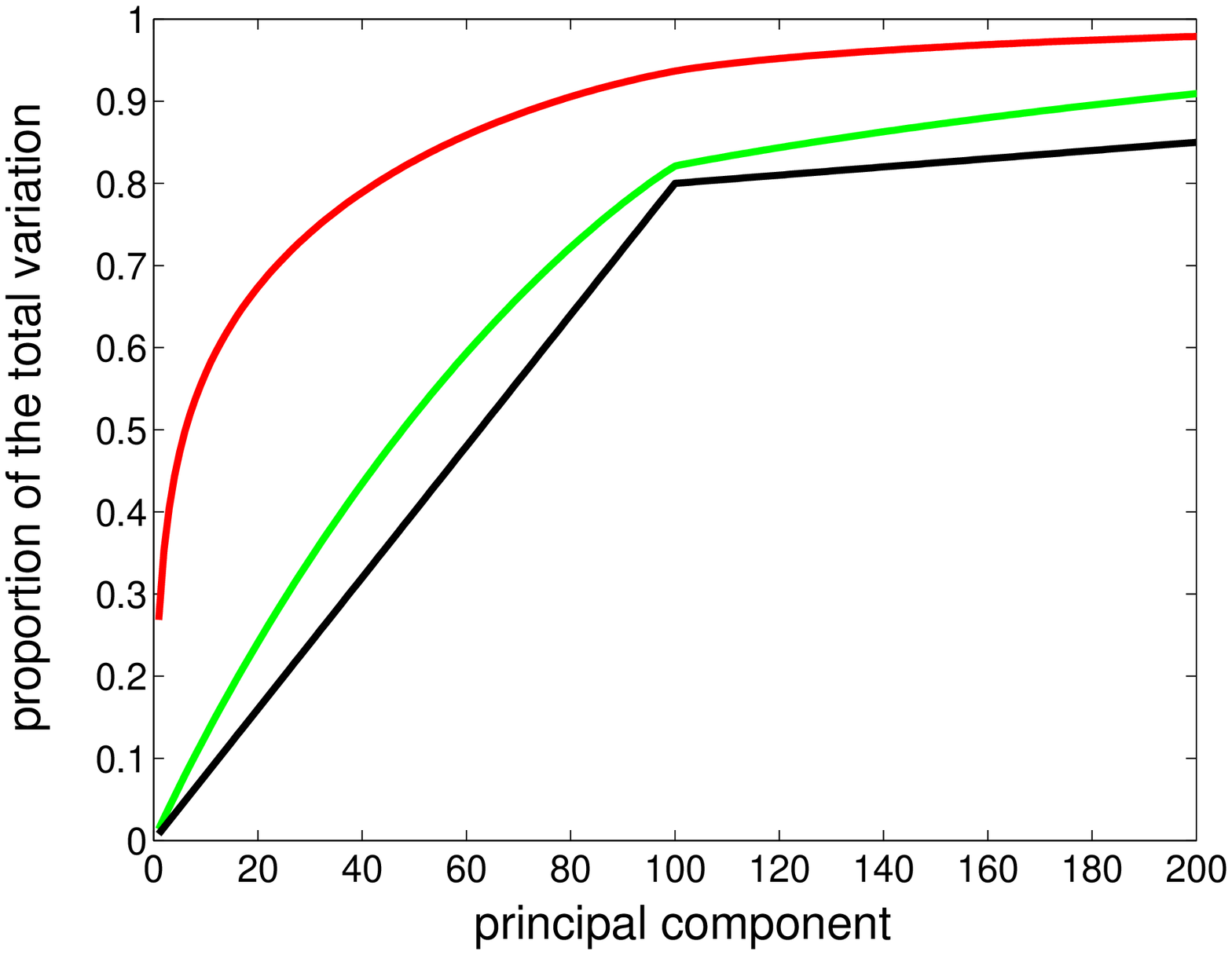}\\[.25cm]
\end{center}
{\bf Fig. 9:} True proportion of the total variation (black line) and proportions obtained by the sample
covariance matrix (red lines) and by the spectral estimator (green lines). The samples are drawn from a
multivariate $t$-distribution with $\nu =\infty$ (i.e. the multivariate normal distribution, upper left),
$\nu=10$ (upper right), $\nu =5$ (lower left), and $\nu =2$ (lower right).\\[.25cm]

In the simulated example of Figure 9 it is assumed that the small eigenvalues are equal. This is equivalent to
the assumption that the residual risk factors are spherically distributed, i.e. that they contain no
more information about the linear dependence structure of $R$. But even if the true eigenvalues are equal
the corresponding estimates will not share this property because of estimation errors. Yet it is important to know whether the residual risk factors have structural information or the differences between the eigenvalue estimates are only caused by random noise. This is not an easy task, especially if the data are not normally distributed and the number of dimensions is large which is the issue of the next section.

\subsection{Signal-Noise Separation}

In the previous section it was mentioned that the central limit theorem fails in the context of high-dimensional data, i.e. if $n/d$ is small. Hence, now we leave the field of classical multivariate analysis and get to the domain of RMT.

Let $\Sigma =\mathcal{ODO}^{\mathrm{T}}\in \R^{d\times d}$ be a spectral decomposition where $\mathcal{D}$ shall be a diagonal matrix containing a `bulk' of small and equal eigenvalues and some large (but
not necessarily equal) eigenvalues. For the sake of simplicity suppose%
\begin{equation*}
\mathcal{D}=\left[
\begin{array}{cc}
cI_{k} & 0 \\
\rule{0cm}{.5cm} 0 & bI_{d-k}%
\end{array}%
\right] \qquad c>b>0,
\end{equation*}%

where $d-k$ is large. Hence $\Sigma$ has two different characteristic manifolds. The `major' one is determined
by the first $k$ column vectors of $\mathcal{O}$ (the `signal part' of $\Sigma$) whereas the `minor' one is
given by the $d-k$ residual column vectors of $\mathcal{O}$ (the `noise part' of $\Sigma$). We are interested in
separating signal from noise that is to say estimating $k$, properly.

For instance, assume that $n=1000$, $d=500$, and that a sample consists of normally distributed random vectors
with covariance matrix $\Sigma$, where $b=1$, $c=5$, and $k=100$. By using the sample covariance matrix and
normalizing the eigenvalues one obtains exemplarily the histogram of eigenvalues given on the left hand of
Figure 10. As might be expected the Mar\v{c}enko-Pastur law is not valid due to the two different regimes of
eigenvalues. In contrast, when focusing on the smallest $400$ eigenvalues, i.e. on the noise part of
$\widehat{\Sigma}$ the Mar\v{c}enko-Pastur law becomes valid as we see on the right hand of Figure 10.

\begin{center}
\includegraphics[scale=.34]{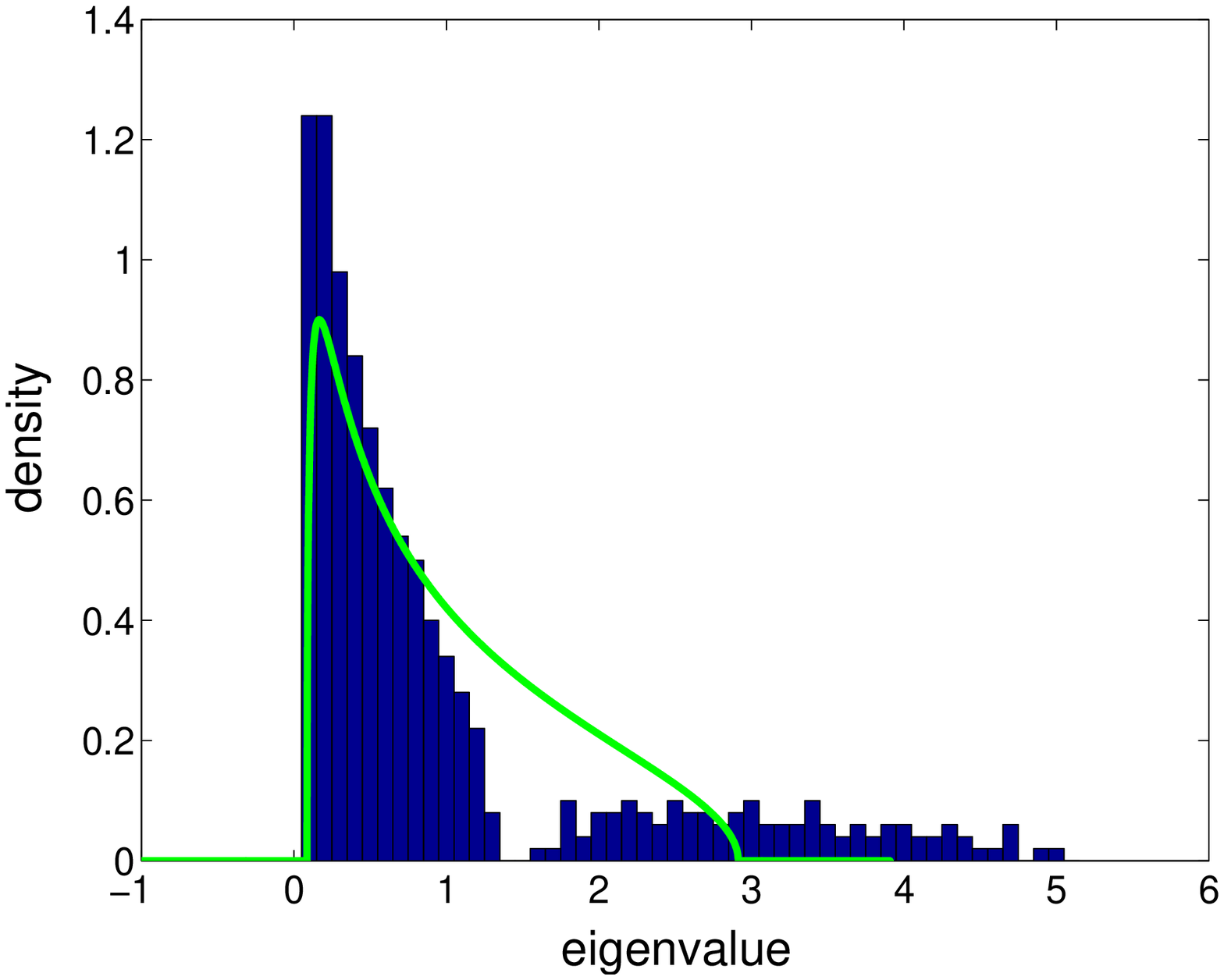}
\includegraphics[scale=.34]{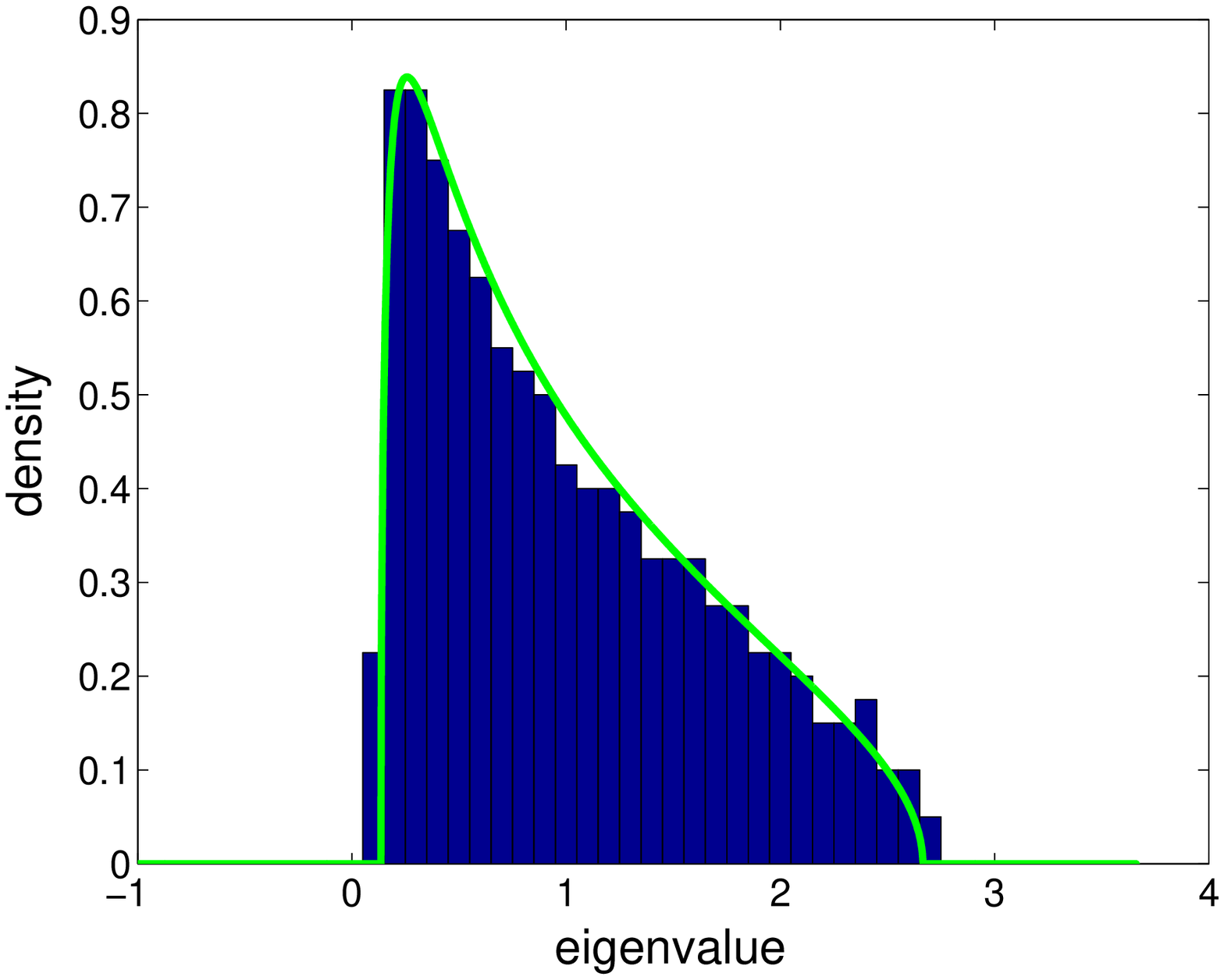}\\[.25cm]
\end{center}
{\bf Fig. 10:} Histogram of all $d=500$ eigenvalues (left hand) and of the noise part (right hand) consisting of
the $d-k=400$ smallest eigenvalues. The Mar\v{c}enko-Pastur law is represented by the green lines.\\[.25cm]

Thus separating signal from noise means sorting out the largest eigenvalues successively until the residual
eigenspectrum is consistent with the Mar\v{c}enko-Pastur law. This is given, e.g., when there are no more
eigenvalues exceeding the Mar\v{c}enko-Pastur upper bound $\lambda _{\max}$. In our case-study this is given
for $397$ eigenvalues (see the figure below), i.e. $\widehat{k}=103$.

\begin{center}
\includegraphics[scale=.35]{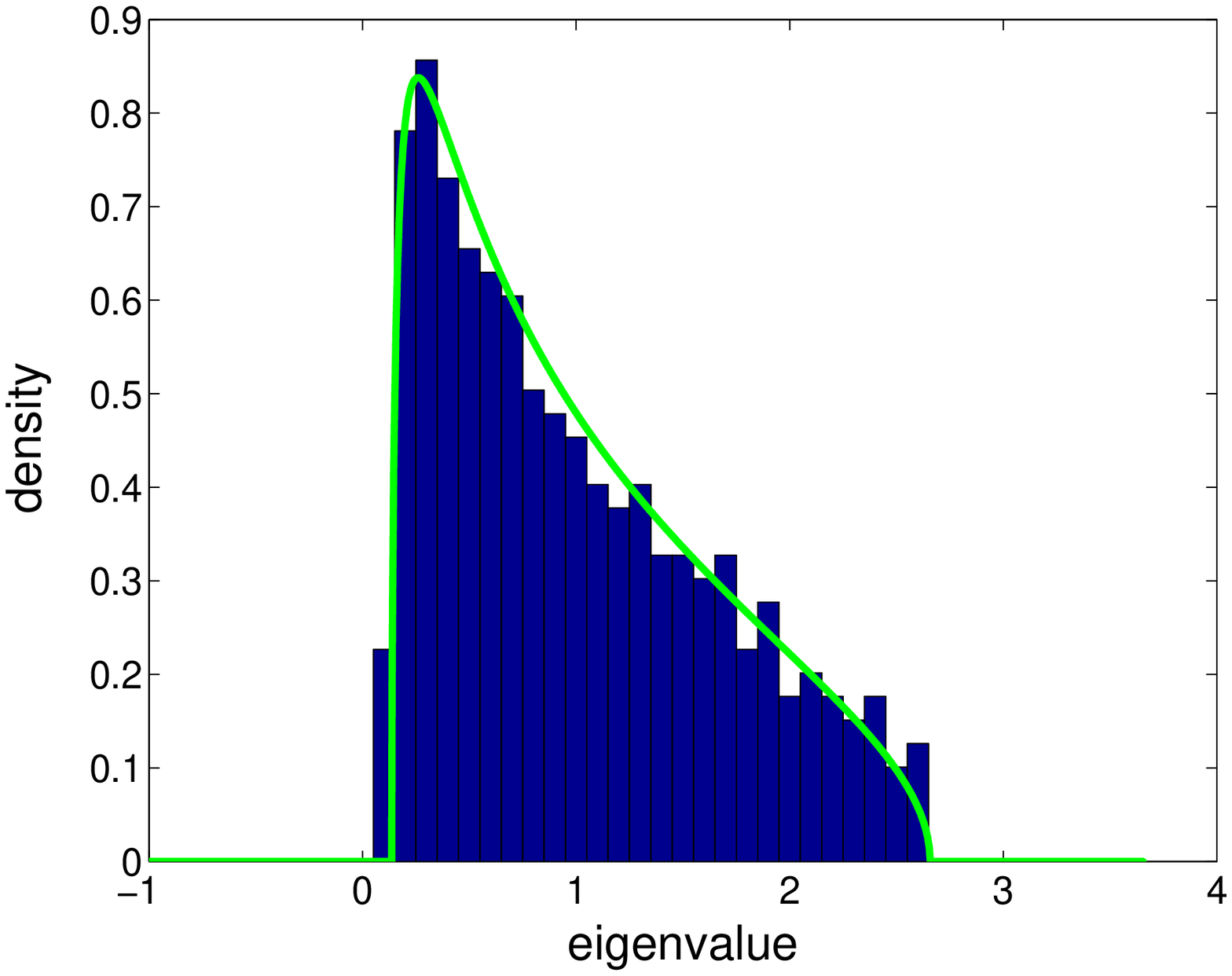}\\[.25cm]
{\bf Fig. 11:} Histogram of the remaining $397$ eigenvalues after signal-noise separation.\\[.25cm]
\end{center}

As it was shown in Section \ref{RMT} this approach is promising only if the data are not regularly varying. Hence
for financial data not the sample covariance matrix but the spectral estimator is proposed for a proper signal-noise
separation.

\section{Conclusions}

Due to the stylized facts of empirical finance the Gaussian distribution hypothesis is not appropriate for the modeling of financial data. For that reason the authors rely on the broad class of generalized elliptical distributions. This class allows for tail dependence and radial asymmetry. Although the sample covariance matrix works quite good with financial data for the most time it is not appropriate for measuring their linear dependence structure. This is due to a few but extreme fluctuations on financial markets.

It is shown that there exists a completely robust ML-estimator (the `spectral estimator') for the dispersion matrix of generalized elliptical distributions. This estimator corresponds to Tyler's M-estimator for elliptical distributions. Further, it is shown that the Mar\v{c}enko-Pastur law fails if the sample covariance matrix is considered as random matrix in the context of elliptically or even generalized elliptically distributed data. This is due to the fact that stochastical independence implies linear independence but conversely uncorrelated random variables are not necessarily independent. In contrast, the Mar\v{c}enko-Pastur law remains valid if the data are uncorrelated and the spectral estimator is considered as random matrix.

The robustness property of the spectral estimator can be demonstrated for several financial applications like, e.g., portfolio risk minimization, principal components analy\-sis, and signal-noise separation. If the data are heavy tailed the principal components analy\-sis tends to underestimate the number of driving risk factors if the sample covariance matrix is used for extracting the eigenspectrum. This means that the contribution of the largest eigenvalues to the total variation of the data is overestimated, systemati\-cally. Consequently, in the context of signal-noise separation the largest eigenvalues are overestimated by the sample covariance matrix. This can be fixed simply by using the spectral estimator, instead.

\end{document}